\documentclass[prl,twocolumn,showpacs,amsmath,amssymb,superscriptaddress]{revtex4-1}
\usepackage{graphicx}
\usepackage{hyperref}
\usepackage{breqn}
\usepackage{xcolor}
\usepackage{color}
\usepackage{booktabs}
\usepackage{float}
\usepackage{longtable}
\usepackage{epsfig}
\usepackage{bm}
\usepackage{dcolumn}

\usepackage{rotating} 

\usepackage[normalem]{ulem}
\usepackage{epstopdf}

\makeatletter
\let\cat@comma@active\@empty
\makeatother
\begin{document}

\title{Chiral magnons in altermagnetic RuO$_2$}

\author{Libor~\v{S}mejkal}
\affiliation{Institut f\"ur Physik, Johannes Gutenberg Universit\"at Mainz, D-55099 Mainz, Germany}
\affiliation{Institute of Physics, Czech Academy of Sciences, Cukrovarnick\'{a} 10, 162 00 Praha 6 Czech Republic}
\author{Alberto~Marmodoro}
\affiliation{Institute of Physics, Czech Academy of Sciences, Cukrovarnick\'{a} 10, 162 00 Praha 6 Czech Republic}
\author{Kyo-Hoon~Ahn}
\affiliation{Institute of Physics, Czech Academy of Sciences, Cukrovarnick\'{a} 10, 162 00 Praha 6 Czech Republic} 
\author{Rafael~Gonz\'{a}lez-Hern\'{a}ndez} 
\affiliation{Grupo de Investigaci\'{o}n en F\'{\i}sica Aplicada, Departamento de F\'{i}sica,Universidad del Norte, Barranquilla, Colombia}
\affiliation{Institut f\"ur Physik, Johannes Gutenberg Universit\"at Mainz, D-55099 Mainz, Germany}
\author{Ilja~Turek}
\affiliation{Institute of Physics of Materials, Czech Academy of Sciences, Zizkova 22, CZ-616 62 Brno, Czech Republic}
\author{Sergiy~Mankovsky}%
\author{Hubert~Ebert}%
\affiliation{Department of Chemistry, Ludwig-Maximilians-University Munich, Butenandtstrasse 11, D-81377 Munich, Germany}
\author{Sunil~W.~D'Souza}
\affiliation{New Technologies-Research Center, University of West Bohemia, Plze\v{n} 3, CZ-30100, Czech Republic}
\author{Ond\v{r}ej~\v{S}ipr}
\affiliation{Institute of Physics, Czech Academy of Sciences, Cukrovarnick\'{a} 10, 162 00 Praha 6 Czech Republic} 
\author{Jairo~Sinova}
\affiliation{Institut f\"ur Physik, Johannes Gutenberg Universit\"at Mainz, D-55099 Mainz, Germany}
\affiliation{Institute of Physics, Czech Academy of Sciences, Cukrovarnick\'{a} 10, 162 00 Praha 6 Czech Republic}
\author{Tom\'a\v{s}~Jungwirth}
\affiliation{Institute of Physics, Czech Academy of Sciences, Cukrovarnick\'{a} 10, 162 00 Praha 6 Czech Republic}
\affiliation{School of Physics and Astronomy, University of Nottingham, Nottingham NG7 2RD, United Kingdom}

\date{\today}

\begin{abstract}
Magnons in ferromagnets have one chirality, and typically are in the GHz range and have  a quadratic dispersion near the zero wavevector. In contrast, magnons in antiferromagnets are commonly considered to have bands with both chiralities that are degenerate across the entire Brillouin zone, and to be in the THz range and to have a linear dispersion near the  center of the Brillouin zone. Here we theoretically demonstrate a new class of magnons on a prototypical  $d$-wave altermagnet RuO$_2$ with the compensated antiparallel magnetic order in the ground state.  Based on density-functional-theory calculations we observe that the THz-range magnon bands in RuO$_2$ have an alternating chirality splitting, 
similar to the alternating spin splitting of the electronic bands, and a linear magnon dispersion near the zero wavevector.  We also show that, overall, the Landau damping of this metallic altermagnet is suppressed due to the spin-split electronic structure, as compared to an artificial antiferromagnetic phase of the same RuO$_2$ crystal with spin-degenerate electronic bands and chirality-degenerate magnon bands. 

\end{abstract}

\maketitle

The study of magnetic excitations offers an important insight into fundamental properties of distinct phases of matter, as well as into  applications of magnetic materials. While incoherent magnons govern the  heat conversion and harvesting phenomena studied in the field of spin-caloritronics \cite{Bauer2012,Mizuguchi2019}, coherent magnons play a key role in the complementary field of magnonics \cite{Neusser2009,Chumak2015a,Sander2017,Pirro2021,Chumak2022}. This research field aims at energy-efficient transmission, storage, and processing of information without the need of charge transport, thus eliminating Joule heating and associated losses. A complete suppression of the motion of both electron and ion charges is an approach towards energy downscaling that is unparalleled in the conventional micro- and opto-electronics \cite{Sander2017,Pirro2021,Chumak2022}. Moreover, compared to photons at the same frequency,  the wavelength of magnons is orders of magnitude shorter which opens a prospect of wave-based information technology scaled down to nanometers. 

Ferromagnetic magnons are chiral and, therefore, can carry spin currents which is the key enabling feature for magnon spintronics \cite{Chumak2015a}. 
However, the ferromagnetic-magnon dispersion is quadratic at low wavevectors which implies a wavevector-dependent group velocity. This hinders a propagation of magnon pulses in the form of stable narrow wave-packets. In addition, ferromagnetic magnonics operates primarily in the GHz range, i.e., at much slower rates than those   offered by photonics. 
 \begin{figure}[h]
	\centering
	\includegraphics[width=0.490 \textwidth]{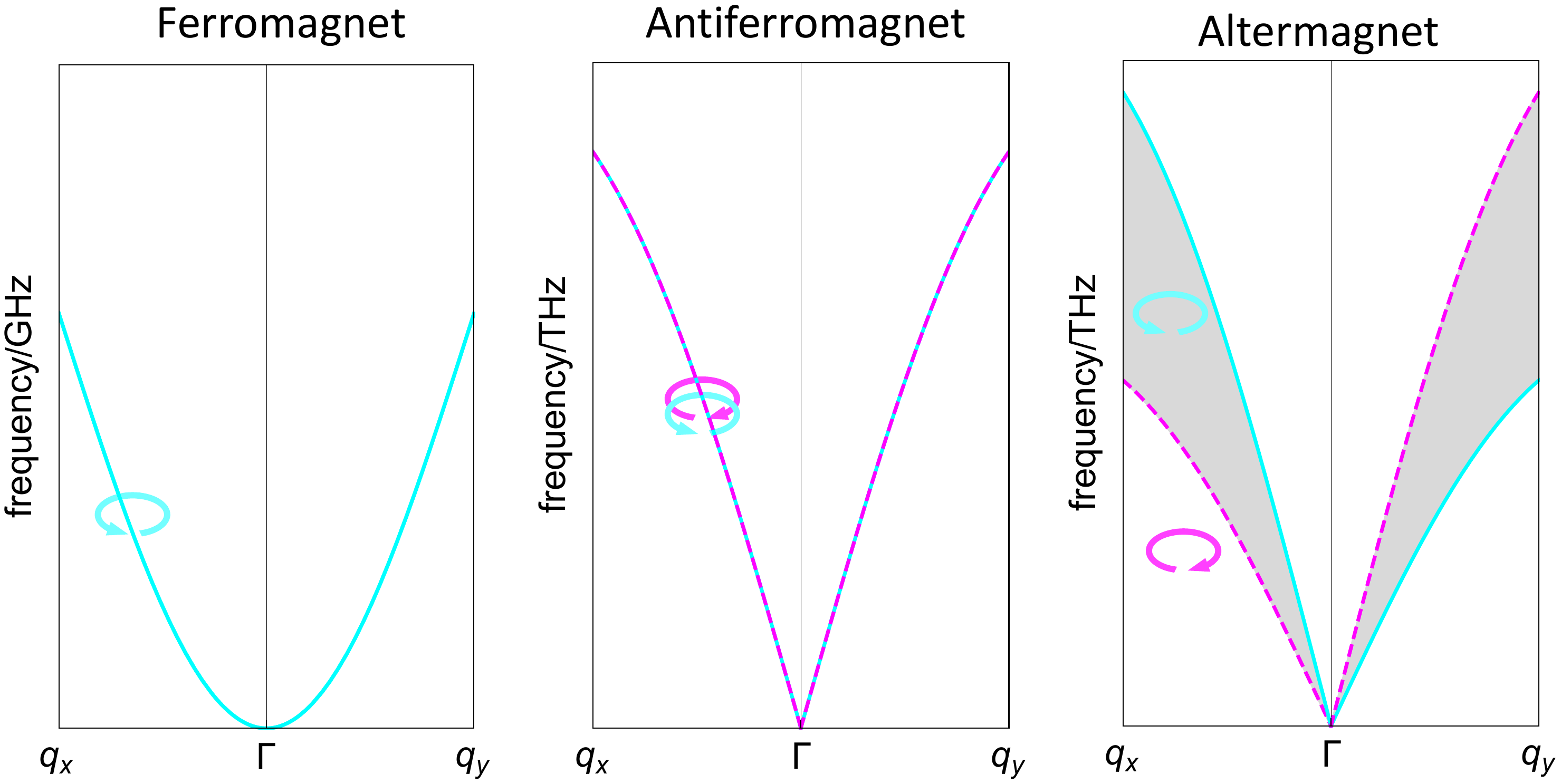} 
	\caption{Schematic representation of the distinct phenomenologies of magnon spectra in a ferromagnet, antiferromagnet and altermagnet.} 
	\label{fig1}
\end{figure}

In contrast, as schematically illustrated in Fig.~\ref{fig1}, antiferromagnetic magnons typically have a photon-like linear dispersion close to the center of the magnonic Brillouin zone when neglecting relativistic effects, and reach the THz frequencies \cite{Gomonay2018,Rezende2019}. However, the opposite-chirality magnon bands  are degenerate across the entire Brillouin zone in the absence of an applied magnetic field. In our Letter we demonstrate that this principal limitation for magnon spintronics can be removed in altermagnets \cite{Smejkal2021a,Smejkal2022a}, while preserving the favorable THz-range and linear magnon dispersion (see Fig.~\ref{fig1}). Confirmed by density-functional-theory (DFT) calculations, we show that in this recently identified third basic magnetic phase \cite{Smejkal2021a,Smejkal2022a},  the magnon bands can have an alternating chirality splitting, similar 
to the alternating spin-splitting of the electronic bands \cite{Smejkal2021a,Smejkal2022a}.

The typically leading contribution to the magnon spectra can be obtained by mapping the spin-dependent electronic structure on the Heisenberg Hamiltonian, $H=-\sum_{i\neq j}J_{ij}\hat{{\bf e}}_i\cdot\hat{{\bf e}}_j$ \cite{Halilov1998,Corticelli2022,Marmodoro2022a}. Here $\hat{{\bf e}}_i$ is the direction of the magnetic moment for
an atom at position ${\bf R}_i$, and  $J_{ij}$ are  Heisenberg exchange coupling parameters. In the Heisenberg model, the
real and spin space transformations are decoupled and  the symmetries of the corresponding magnon bands can be described by the non-relativistic spin-group formalism \cite{Brinkman1966,Corticelli2022}.

In antiferromagnets, crystallographic translation or inversion symmetry transformations connecting  opposite-spin sublattices protect the degeneracy of the magnon bands with the opposite chirality \cite{Brinkman1966,Corticelli2022}. Remarkably, this has been commonly illustrated on the opposite-spin sublattices of rutile crystals, while omitting the presence of non-magnetic  atoms  in these crystals \cite{Brinkman1966,Rezende2019,Corticelli2022}. 
However, as highlighted in recent theoretical studies of the electronic structure of the metallic rutile RuO$_2$, the non-magnetic O-atoms  break the translation and inversion symmetries connecting the opposite-spin sublattices \cite{Smejkal2020,Smejkal2021a,Smejkal2022a}.  
The remaining symmetries connecting the opposite-spin sublattices, that protect the zero net magnetization in the non-relativistic limit, are crystallographic (proper or improper) rotations \cite{Smejkal2021a,Smejkal2022a}. 
The symmetries of RuO$_2$ were predicted to allow for the relativistic anomalous Hall effect \cite{Smejkal2020,Feng2020a,Smejkal2021b},  and for non-relativistic spin currents \cite{Gonzalez-Hernandez2021},  giant and tunneling magnetoresitance \cite{Smejkal2022,Shao2021}, and spin-split band structure \cite{Smejkal2021a,Ahn2019}. Instead of classical antiferromagnetism \cite{Neel1953,Brinkman1966,Rezende2019,Corticelli2022},  the rutiles, including RuO$_2$, are the prototypical representatives of altermagnetism according to the recent spin-group symmetry classification of collinear magnetic phases \cite{Smejkal2021a}.  In general, altermagnets exhibit characteristic  $d$, $g$, or $i$-wave anisotropy and alternating spin polarization across the band-structure Brillouin zone. This follows from their spin-group symmetry that is exclusively distinct from the spin-group symmetries of conventional ferromagnetism and antiferromagnetism  \cite{Smejkal2021a,Smejkal2022a}. 
The theoretical predictions of the $d$-wave altermagnetic phenomenolgy in RuO$_2$ have been supported by recent measurements of the anomalous Hall effect \cite{Feng2020a} and spin currents in RuO$_2$ \cite{Bose2022,Bai2021,Karube2022}.
Anomalous Hall effect was also reported and predicted to arise in other altermagnetic candidates \cite{Samanta2020,Naka2020,Reichlova2020,Mazin2021,Betancourt2021,Naka2022}.

In the first part of this Letter, we map the DFT electronic band structure on the Heisenberg Hamiltonian and obtain the chirality-split, as well as the chirality-degenerate, parts of the spectrum of the undamped magnons across the RuO$_2$ Brillouin zone. In the second part, we include into the consideration also  the spectrum of Stoner excitations, which is expected to provide the leading mechanism limiting the lifetime of magnons \cite{Costa2003,Marmodoro2022a} in metallic RuO$_2$. On one hand, the spin-split parts of the electronic band structure are expected to suppress this Landau damping, in analogy to ferromagnets \cite{Yosida1996,Costa2003,Kakehashi2013,Marmodoro2022a}. On the other hand, the altermagnetic spin-group symmetries also protect the presence of spin-degenerate nodal surfaces crossing the $\boldsymbol\Gamma$-point \cite{Smejkal2021a,Smejkal2022a}, and these can be expected to generate an antiferromagnetic-like enhancement of the Landau damping. We will qualitatively estimate the strength of the Landau damping in the altermagnetic phase of RuO$_2$  by comparing it to the calculated magnon lifetime in an artificial antiferromagnetic phase of the same RuO$_2$ crystal with spin-degenerate electronic bands and chirality-degenerate magnonic bands across the entire Brillouin zone.

 \begin{figure}[th]
	\centering
	\includegraphics[width=0.490 \textwidth]{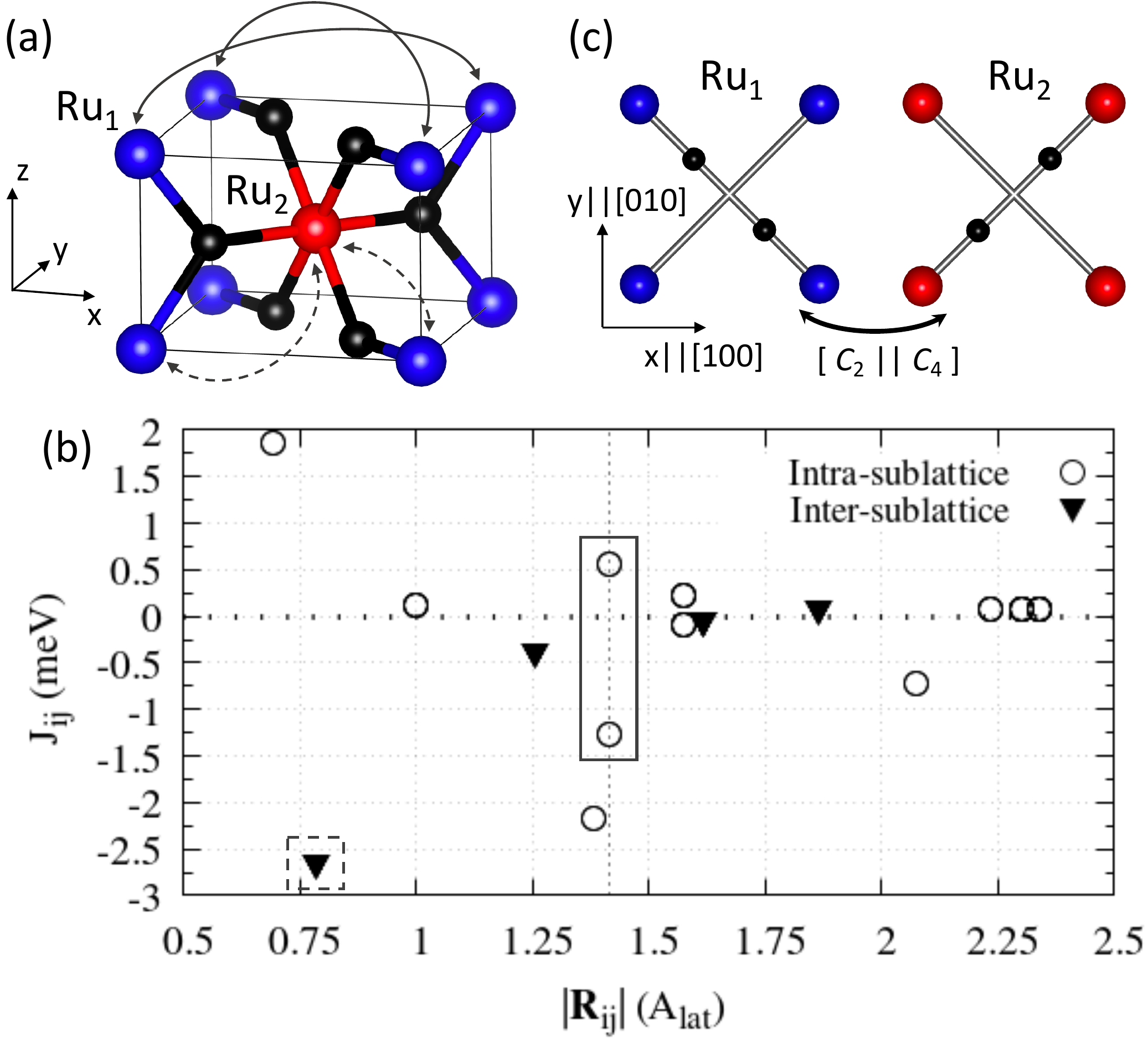} 
	\caption{(a) Model of the crystal structure of the altermagnetic rutile phase of RuO$_2$. The opposite-spin sublattices are labeled by Ru$_1$ and Ru$_2$ and further distinguished by the blue and red color. The black spheres correspond to O atoms. Inter/intra-sublattice exchange interactions between pairs of Ru atoms, discussed in the text, are highlighted  by dashed/solid double-arrow curves. (b) DFT Heisenberg exchange parameters as a function of the separation of Ru atoms. The dashed/solid rectangle highlights the exchange parameters corresponding to the dashed/solid double-arrow curves in panel (a). (c) The opposite spin sublattices with the highlighted spin-group symmetry of the altermagnetic RuO$_2$, which combines a two-fold spin-space rotation $C_{2}$ with a four-fold crystallographic-space rotation $C_{4}$.} 
	\label{fig2}
\end{figure}

 \begin{figure}[h]
	\includegraphics[width=0.4 \textwidth]{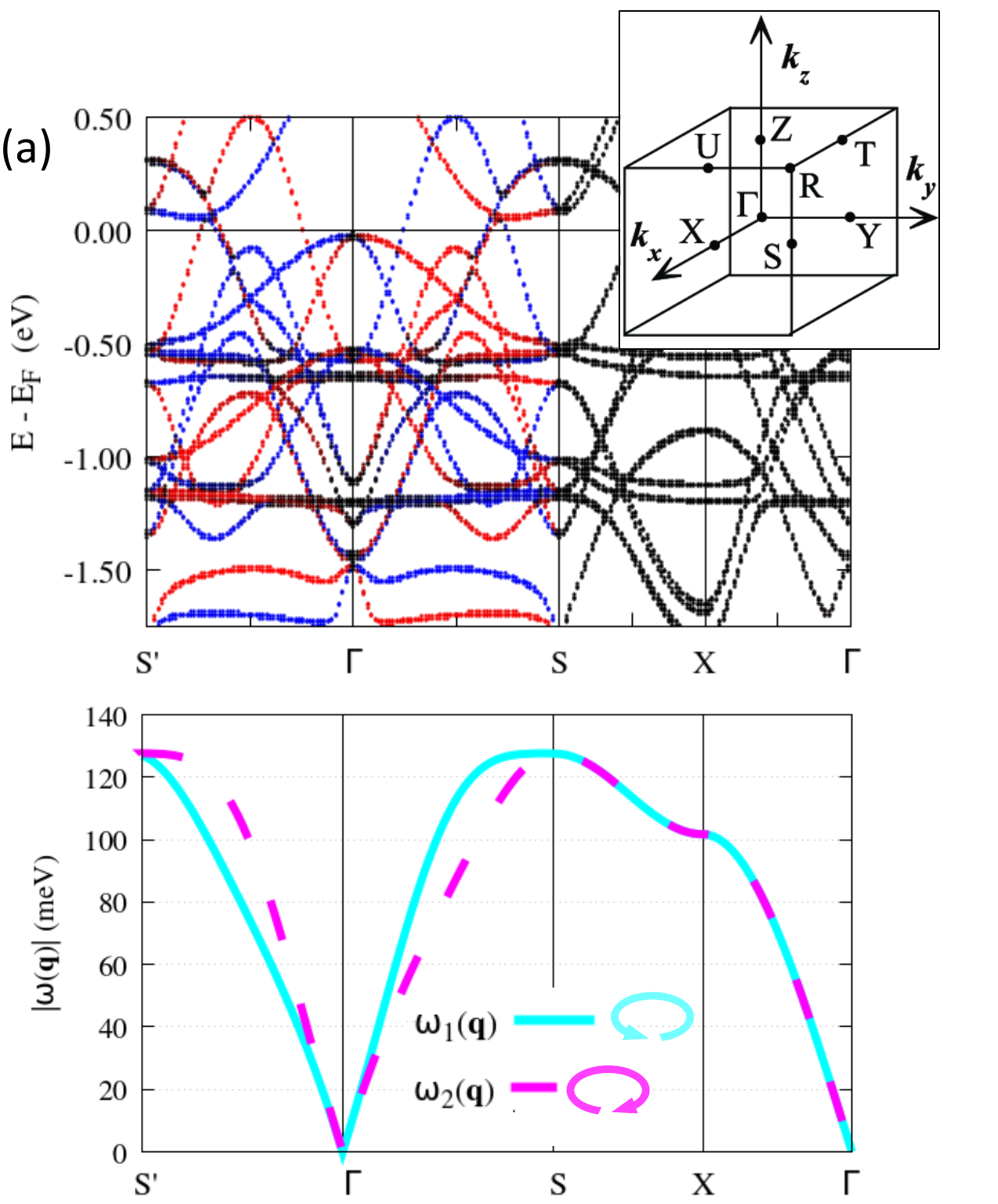} 
         \hspace*{-0.1cm}
         \includegraphics[width=0.4\textwidth]{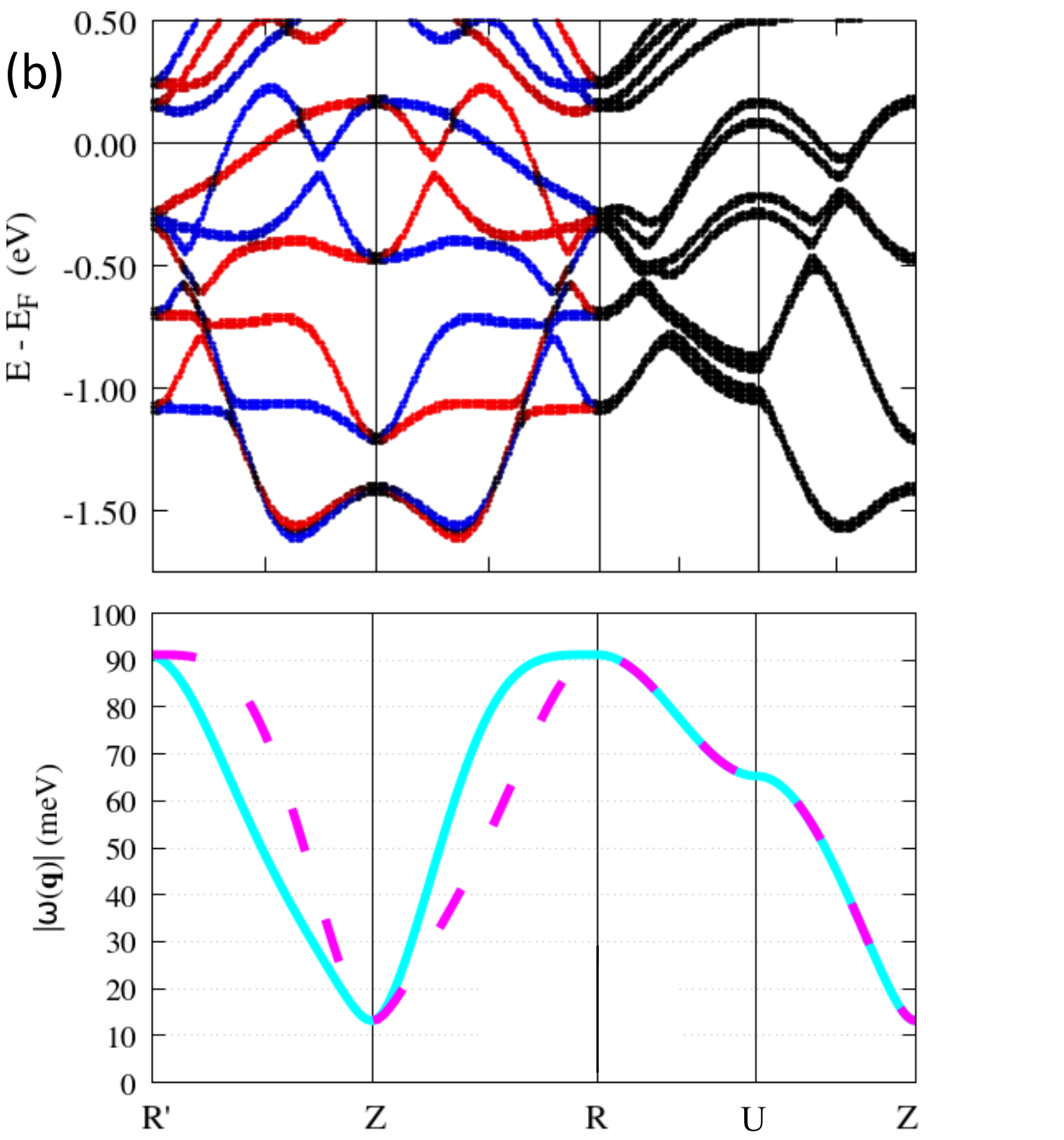} 
	\caption{(a) Top: Electronic band structure of the altermagnetic phase of RuO$_2$ plotted along a Brillouin zone path in a $k_x-k_y$ plane crossing the $\boldsymbol\Gamma$-point. Red and blue colors mark projections on opposite-spin channels. Bottom: Corresponding magnonic band structure. Purple and cyan colors mark opposite magnon chiralities. (b) Same as (a) plotted along a Brillouin zone path in a $k_x-k_y$ plane crossing the {\bf Z}-point.} 
	\label{fig3}
\end{figure}

In the DFT electronic-structure calculations \cite{Ebert2011a}, we consider the experimental lattice parameters of RuO$_2$, taken from Ref.~\cite{Haines1997} (for more details see Supplementary information). The magnon spectra of RuO$_2$ were produced by mapping the DFT results onto the Heisenberg Hamiltonian and by diagonalizing the associated Landau-Lifschitz equation of motion for the adiabatic magnon dispersion.  The coupling parameters were determined using the magnetic force theorem, as detailed in Ref.~\cite{Mankovsky2017}.
For the DFT-based calculation of the Landau damping we adopt  the approximation of Ref.~\cite{Marmodoro2022a}. (For more details see Supplementary information.)

In Fig.~\ref{fig2}a we show the rutile crystal structure of RuO$_{\text{2}}$.  
The calculated pairwise Heisenberg exchange interaction parameters among different sets of Ru atoms mutually separated by $|{\bf R}_{ij}|$ are shown in  Fig~\ref{fig2}b.  Among those, we observe that all inter-sublattice exchange interactions are negative or negligibly small. The resulting compensated antiparallel ordering is consistent with experimental neutron scattering data on RuO$_{\text{2}}$ \cite{Berlijn2017a}.  The nearest-neighbor inter-sublattice exchange parameter is highlighted in Fig~\ref{fig2}b by a dashed rectangle and corresponds to the pairs of Ru atoms connected by dashed double-arrow curves in Fig.~\ref{fig2}a.

Next we analyze the intra-sublattice exchange parameters. Among those, the ones corresponding to Ru atoms separated by $\sqrt{2}$ $A_{lat}$ play a decisive role in the resulting alternating chirality-splitting of the magnon spectra, plotted in Fig.~\ref{fig3} side-by-side the spin-split electronic bands.  ($A_{lat}$ denotes the  $x-y$ plane lattice constant, see Supplementary information.) These ${J}_{ij}$'s are highlighted in Fig.~\ref{fig2}b by a solid rectangle and correspond to the pairs of Ru atoms connected by solid double-arrow curves in Fig~\ref{fig2}a. They are sizable in magnitude when compared to the first and second-nearest-neighbour exchange parameters (owing to the itinerant origin of magnetism in RuO$_{2}$) shown in Fig.~\ref{fig2}b.  However, unlike the latter, the intra-sublattice  ${J}_{ij}$'s  for Ru atoms separated by $|{\bf R}_{ij}|=\sqrt{2}$ $A_{lat}$ depend on the direction of ${\bf R}_{ij}$. They have a different magnitude and even opposite sign, depending on whether the exchange interaction acts along the $y=-x$ or along the $90^{\circ}$-rotated $y=+x$ diagonal.  
We attribute the strong dependence of these ${J}_{ij}$ parameters on the direction of ${\bf R}_{ij}$ to the presence or absence of the O atoms along the corresponding bonds, as seen in Fig.~\ref{fig2}a. Furthermore,  these anisotropies (i.e. the presence/absence of O atoms along the bonds) on the two opposite-spin Ru sublattices are related by the altermagnetic spin-group symmetry $\left[C_{2} \vert\vert C_{4} \right]$, which combines a two-fold spin-space rotation with a four-fold crystallographic-space rotation \cite{Smejkal2021a}. This spin-group symmetry is important for understanding the chirality-split magnonic spectrum in RuO$_{2}$.

Fig.~\ref{fig3} shows the electronic and magnonic band dispersions along Brillouin zone lines in the $k_x-k_y$ planes crossing the $\boldsymbol\Gamma$-point and the {\bf Z}-point. There are two magnon eigenmodes, $\omega_n({\bf q})$, reflecting the number of magnetic sublattices. The corresponding eigenfrequency of the wave-like angular precession has opposite sign along the two branches, i.e., it describes clockwise or anti-clockwise rotation of the Ru atomic moments around the $\hat{z}$-axis (the easy-axis of the magnetic moments in RuO$_2$ \cite{Berlijn2017a,Smejkal2020,Feng2020a}). In either case, the excitation of magnons requires to supply energy into the system, proportional to $|\omega_n({\bf q})|$.

In the Supplementary information, we highlight the magnitude of the corresponding magnon eigenvector components, which indicates how much each of the  two opposite-spin Ru sublattices have their moments tilted away from the $\hat{z}$-direction, i.e., from the ground state order \cite{Halilov1998}. This representation shows how the excitations involve the opposite-spin sublattices in different fashion as a function of the wavevector {\bf q} within the Brillouin zone. Near the $\boldsymbol\Gamma$-point, the collective precession affects to the same extent the magnetic moments of both Ru sublattices. In contrast, towards the boundary of the Brillouin zone, each eigenmode is dominantly carried by one or the other Ru sublattice, respectively.

The key observation in Fig.~\ref{fig3} is that the opposite-chirality magnonic bands can be split, as illustrated by calculations along the $\boldsymbol\Gamma - {\bf S}$ and $ {\bf Z} - {\bf R}$ paths. The chirality splitting of the magnonic bands is similar 
to the spin-splitting of the electronic bands along the same paths (top panels in Figs.~\ref{fig3}a,b). Magnon excitations of one chirality eigenmode occur at a larger energy cost than for the opposite chirality. The order gets reversed along the orthogonal directions  $\boldsymbol\Gamma - {\bf S}^\prime$ and $ {\bf Z} - {\bf R}^\prime$. This is again similar
to the alternating sign of the spin splitting observed in the electronic band structure. The alternating sign of the chirality splitting and spin splitting imposes that  the magnonic and electronic bands are spin degenerate in parts of the Brillouin zone. In Fig.~\ref{fig3}, this is illustrated along the ${\bf S} - {\bf X} - \boldsymbol\Gamma$ and ${\bf R} - {\bf U} - {\bf Z}$ paths.

We can numerically track the origin of the chirality splitting to the directionality dependence in the Heisenberg exchange interaction
discussed above, accounting for the contribution of the exchange couplings within spheres of different radius. For $|{\bf R}_{ij}| < \sqrt{2} A_{lat}$,
the  $J_{ij}$ parameters depend only on the distance (see Fig.~\ref{fig2}b). Calculations of the magnon dispersion truncating the lattice Fourier transformation to this shorter cutoff produce no lifting of the degeneracy of the opposite-chirality magnon bands. On the other hand, the anisotropy of some of the exchange interactions (e.g. those highlighted in Fig.~\ref{fig2}a by dashed double-arrow curves)
generates the chirality splitting of the magnon eigenmodes in the RuO$_2$ altermagnet. 

We emphasize that the anisotropy of the exchange interactions alone is not sufficient for generating a chirality-split magnon spectrum. The anisotropy has to be accompanied by the spin-group symmetries of the altermagnetic phase of RuO$_2$. To highlight this point, we have performed reference calculations for an artificial antiferromagnetic phase of RuO$_2$, constructed from the altermagnetic phase by doubling the unit cell along the $z$-axis, which introduces an additional translation symmetry connecting the opposite-spin sublattices (see Supplementary information). Despite the persisting anisotropy of some of the $J_{ij}$'s, the spin-group symmetry of this antiferromagnetic phase protects the spin-degeneracy of the electronic bands across the whole Brillouin zone  \cite{Smejkal2021a,Smejkal2022a}. Our DFT calculations shown in the Supplementary information confirm the corresponding chirality-degeneracy of the magnon bands.

Another important observation in Fig.~\ref{fig3} is the linear energy dependence $|\omega_n({\bf q})| \propto |{\bf q}|$ 
in the ${\bf q} \to \boldsymbol\Gamma$ limit, and the THz range of the altermagnetic magnons in RuO$_2$. This is analogous to the conventional magnon dispersions in antiferromagnets \cite{Gomonay2018,Rezende2019}, and contrasts with  the quadratic dispersion and GHz range typical of ferromagnetic magnons.

 \begin{figure}[h]
	\centering
	\includegraphics[width=0.470 \textwidth]{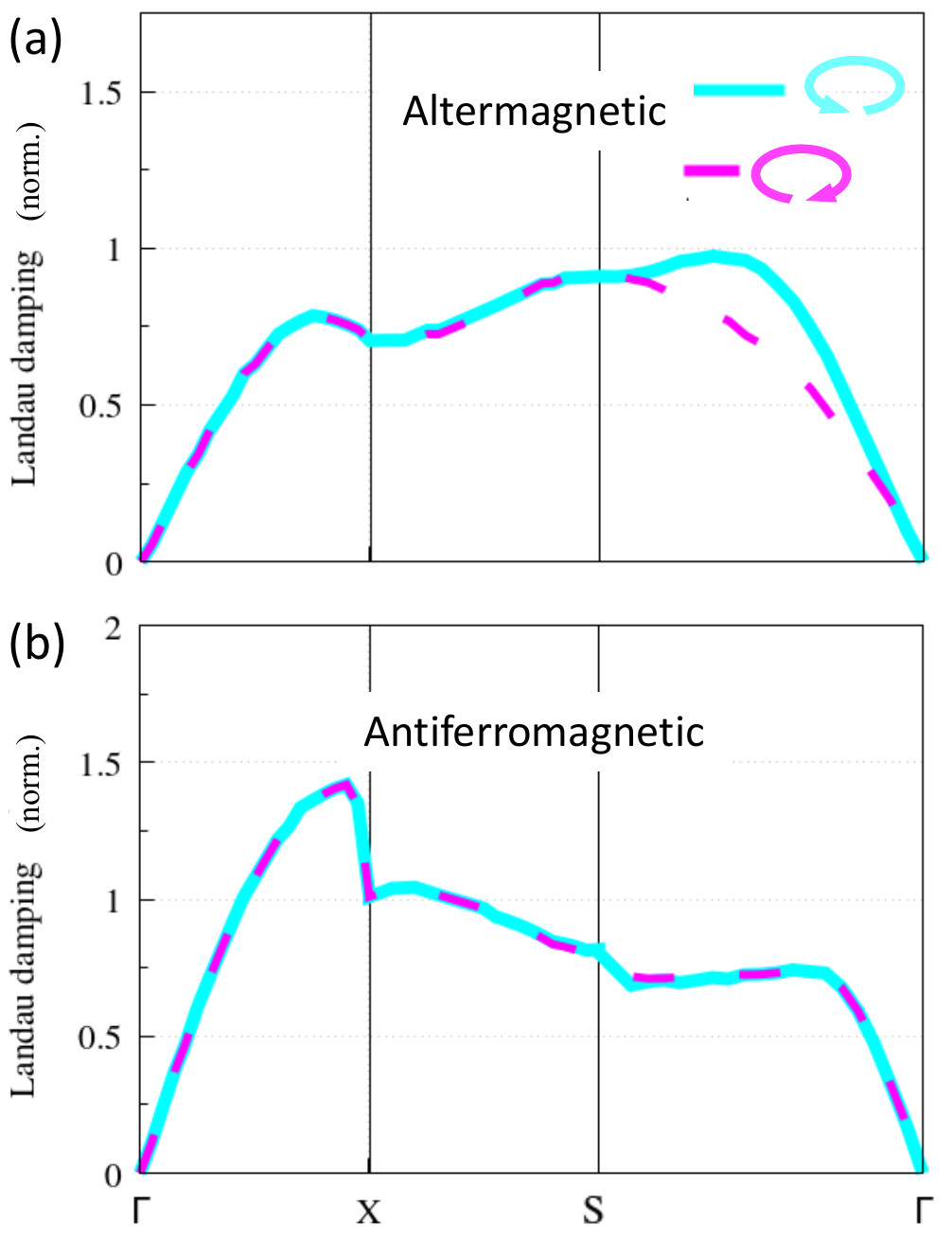} 
	\caption{Comparison of the Landau damping in the altermagnetic (a) and artificial antiferromagnetic (b) phase of RuO$_2$, normalized to the maximum Landau damping in the altermagnetic phase for the depicted path in the Brillouin zone. Purple and cyan colors mark opposite magnon chiralities} 
	\label{fig4}
\end{figure}

We now move on to the discussion of the Landau damping, determined by the intensity of Stoner single-particle spin-flip transitions. The approximated Landau damping calculated by the method described in Ref.~\cite{Marmodoro2022a} and in the  Supplementary information is shown in Fig.~\ref{fig4}. The intensity of Stoner single-particle spin-flip transitions is degenerate along the ${\bf S} - {\bf X} - \boldsymbol\Gamma$ path, or non-degenerate along the $\boldsymbol\Gamma - {\bf S}$  path, in the same way as observed for the magnon dispersion $\omega_n({\bf q})$ in Figs.~\ref{fig3}. In Figs.~\ref{fig4} we qualitatively compare the Landau damping calculated along the above Brillouin zone paths for the altermagnetic phase of RuO$_2$ and for the artificial antiferromagnetic phase. In both panels, the data are normalized to the maximum value of the Landau damping in the altermagnetic phase for the depicted path in the Brillouin zone. As expected, the antiferromagnetic phase leads to the degeneracy in the Landau damping of the opposite-chirality magnon eigenmodes. As also anticipated from the spin degenerate band structure over the entire Brillouin zone, the Landau damping tends to be, overall, stronger in the antiferromagnetic phase.

Near the $\boldsymbol\Gamma$ point, the calculated Landau damping in the altermagnetic phase of RuO$_2$ shows a non-zero intensity of the competing Stoner excitations, which is analogous to the behavior of metallic antiferromnagnets \cite{Yosida1996}. In comparison, metallic ferromagnets present a Stoner gap, i.e., no Landau damping for ${\bf q} \to \boldsymbol\Gamma$. This is a consequence of the splitting of the electronic bands  into majority and minority-spin states. Single-particle spin-flip transitions  that can compete with the collective magnetic excitations are only possible upon overcoming the above gap \cite{Yosida1996}.
In the altermagnetic phase of RuO$_2$, the electronic bands over most the Brillouin-zone volume are also spin split. This explains the observed overall suppression of the Landau damping, as compared to the artificial antiferromagnetic phase. The Landau damping in altermagnetic RuO$_2$, therefore, shows features that are intermediate between the antiferromagnetic and ferromagnetic cases. 
 
To conclude, we have demonstrated that similarly 
to the electronic band structure, the degeneracy of  magnon bands with opposite chirality is  lifted in altermagnetic RuO$_2$, with the sign of the chirality-splitting alternating across the magnon Brillouin zone. This indicates that, on one hand, magnons in altermagnets can be chiral and carry spin currents, similar to ferromagnets but with highly anisotropic characteristics. On the other hand, the dispersion of altermagnetic magnons can be linear around the $\boldsymbol\Gamma$-point, similar to antiferromagnets.  These unconventional characteristics make altermagnets a promising material platform for magnonics research, including the exploration  of  the emission, propagation and detection of ultra-short (ps-scale) pulses of chiral THz magnons of wavelengths orders of magnitude smaller than the wavelengths of the THz photons.

\section*{Acknowledgments}
We acknowledge funding from the Czech Science Foundation Grant No. 19-18623X and 19-06433S,  the Ministry of Education of the
Czech Republic Grants No. LM2018096, LM2018110, LM2018140, and LNSM-LNSpin, and  the EU FET Open RIA Grant No. 766566, Deutsche Forschungsgemeinschaft Grant TRR 173 268565370 (project A03),   the computing time supported by the project
CEDAMNF CZ.02.1.01/0.0/0.0/15\_003/0000358 (Ministry of Education,Youth and Sport) and by the "Information Technology for Innovation" (IT4I) grant OPEN-22-42, and granted on the supercomputer Mogon at Johannes Gutenberg University Mainz (hpc.uni-mainz.de) and the support of Alexander Von Humboldt Foundation.


\begin{thebibliography}{42}%
\makeatletter
\providecommand \@ifxundefined [1]{%
 \@ifx{#1\undefined}
}%
\providecommand \@ifnum [1]{%
 \ifnum #1\expandafter \@firstoftwo
 \else \expandafter \@secondoftwo
 \fi
}%
\providecommand \@ifx [1]{%
 \ifx #1\expandafter \@firstoftwo
 \else \expandafter \@secondoftwo
 \fi
}%
\providecommand \natexlab [1]{#1}%
\providecommand \enquote  [1]{``#1''}%
\providecommand \bibnamefont  [1]{#1}%
\providecommand \bibfnamefont [1]{#1}%
\providecommand \citenamefont [1]{#1}%
\providecommand \href@noop [0]{\@secondoftwo}%
\providecommand \href [0]{\begingroup \@sanitize@url \@href}%
\providecommand \@href[1]{\@@startlink{#1}\@@href}%
\providecommand \@@href[1]{\endgroup#1\@@endlink}%
\providecommand \@sanitize@url [0]{\catcode `\\12\catcode `\$12\catcode
  `\&12\catcode `\#12\catcode `\^12\catcode `\_12\catcode `\%12\relax}%
\providecommand \@@startlink[1]{}%
\providecommand \@@endlink[0]{}%
\providecommand \url  [0]{\begingroup\@sanitize@url \@url }%
\providecommand \@url [1]{\endgroup\@href {#1}{\urlprefix }}%
\providecommand \urlprefix  [0]{URL }%
\providecommand \Eprint [0]{\href }%
\providecommand \doibase [0]{http://dx.doi.org/}%
\providecommand \selectlanguage [0]{\@gobble}%
\providecommand \bibinfo  [0]{\@secondoftwo}%
\providecommand \bibfield  [0]{\@secondoftwo}%
\providecommand \translation [1]{[#1]}%
\providecommand \BibitemOpen [0]{}%
\providecommand \bibitemStop [0]{}%
\providecommand \bibitemNoStop [0]{.\EOS\space}%
\providecommand \EOS [0]{\spacefactor3000\relax}%
\providecommand \BibitemShut  [1]{\csname bibitem#1\endcsname}%
\let\auto@bib@innerbib\@empty
\bibitem [{\citenamefont {Bauer}\ \emph {et~al.}(2012)\citenamefont {Bauer},
  \citenamefont {Saitoh},\ and\ \citenamefont {van Wees}}]{Bauer2012}%
  \BibitemOpen
  \bibfield  {author} {\bibinfo {author} {\bibfnamefont {G.~E.~W.}\
  \bibnamefont {Bauer}}, \bibinfo {author} {\bibfnamefont {E.}~\bibnamefont
  {Saitoh}}, \ and\ \bibinfo {author} {\bibfnamefont {B.~J.}\ \bibnamefont {van
  Wees}},\ }\href {\doibase 10.1038/nmat3301} {\bibfield  {journal} {\bibinfo
  {journal} {Nature Materials}\ }\textbf {\bibinfo {volume} {11}},\ \bibinfo
  {pages} {391} (\bibinfo {year} {2012})}\BibitemShut {NoStop}%
\bibitem [{\citenamefont {Mizuguchi}\ and\ \citenamefont
  {Nakatsuji}(2019)}]{Mizuguchi2019}%
  \BibitemOpen
  \bibfield  {author} {\bibinfo {author} {\bibfnamefont {M.}~\bibnamefont
  {Mizuguchi}}\ and\ \bibinfo {author} {\bibfnamefont {S.}~\bibnamefont
  {Nakatsuji}},\ }\href {\doibase 10.1080/14686996.2019.1585143} {\bibfield
  {journal} {\bibinfo  {journal} {Science and Technology of Advanced
  Materials}\ }\textbf {\bibinfo {volume} {20}},\ \bibinfo {pages} {262}
  (\bibinfo {year} {2019})}\BibitemShut {NoStop}%
\bibitem [{\citenamefont {Neusser}\ and\ \citenamefont
  {Grundler}(2009)}]{Neusser2009}%
  \BibitemOpen
  \bibfield  {author} {\bibinfo {author} {\bibfnamefont {S.}~\bibnamefont
  {Neusser}}\ and\ \bibinfo {author} {\bibfnamefont {D.}~\bibnamefont
  {Grundler}},\ }\href {\doibase 10.1002/adma.200900809} {\bibfield  {journal}
  {\bibinfo  {journal} {Advanced Materials}\ }\textbf {\bibinfo {volume}
  {21}},\ \bibinfo {pages} {2927} (\bibinfo {year} {2009})}\BibitemShut
  {NoStop}%
\bibitem [{\citenamefont {Chumak}\ \emph {et~al.}(2015)\citenamefont {Chumak},
  \citenamefont {Vasyuchka}, \citenamefont {Serga},\ and\ \citenamefont
  {Hillebrands}}]{Chumak2015a}%
  \BibitemOpen
  \bibfield  {author} {\bibinfo {author} {\bibfnamefont {A.~V.}\ \bibnamefont
  {Chumak}}, \bibinfo {author} {\bibfnamefont {V.~I.}\ \bibnamefont
  {Vasyuchka}}, \bibinfo {author} {\bibfnamefont {A.~A.}\ \bibnamefont
  {Serga}}, \ and\ \bibinfo {author} {\bibfnamefont {B.}~\bibnamefont
  {Hillebrands}},\ }\href {\doibase 10.1038/nphys3347} {\bibfield  {journal}
  {\bibinfo  {journal} {Nature Physics}\ }\textbf {\bibinfo {volume} {11}},\
  \bibinfo {pages} {453} (\bibinfo {year} {2015})}\BibitemShut {NoStop}%
\bibitem [{\citenamefont {Sander}\ \emph {et~al.}(2017)\citenamefont {Sander},
  \citenamefont {Valenzuela}, \citenamefont {Makarov}, \citenamefont {Marrows},
  \citenamefont {Fullerton}, \citenamefont {Fischer}, \citenamefont {McCord},
  \citenamefont {Vavassori}, \citenamefont {Mangin}, \citenamefont {Pirro},
  \citenamefont {Hillebrands}, \citenamefont {Kent}, \citenamefont {Jungwirth},
  \citenamefont {Gutfleisch}, \citenamefont {Kim},\ and\ \citenamefont
  {Berger}}]{Sander2017}%
  \BibitemOpen
  \bibfield  {author} {\bibinfo {author} {\bibfnamefont {D.}~\bibnamefont
  {Sander}}, \bibinfo {author} {\bibfnamefont {S.~O.}\ \bibnamefont
  {Valenzuela}}, \bibinfo {author} {\bibfnamefont {D.}~\bibnamefont {Makarov}},
  \bibinfo {author} {\bibfnamefont {C.~H.}\ \bibnamefont {Marrows}}, \bibinfo
  {author} {\bibfnamefont {E.~E.}\ \bibnamefont {Fullerton}}, \bibinfo {author}
  {\bibfnamefont {P.}~\bibnamefont {Fischer}}, \bibinfo {author} {\bibfnamefont
  {J.}~\bibnamefont {McCord}}, \bibinfo {author} {\bibfnamefont
  {P.}~\bibnamefont {Vavassori}}, \bibinfo {author} {\bibfnamefont
  {S.}~\bibnamefont {Mangin}}, \bibinfo {author} {\bibfnamefont
  {P.}~\bibnamefont {Pirro}}, \bibinfo {author} {\bibfnamefont
  {B.}~\bibnamefont {Hillebrands}}, \bibinfo {author} {\bibfnamefont {A.~D.}\
  \bibnamefont {Kent}}, \bibinfo {author} {\bibfnamefont {T.}~\bibnamefont
  {Jungwirth}}, \bibinfo {author} {\bibfnamefont {O.}~\bibnamefont
  {Gutfleisch}}, \bibinfo {author} {\bibfnamefont {C.~G.}\ \bibnamefont {Kim}},
  \ and\ \bibinfo {author} {\bibfnamefont {A.}~\bibnamefont {Berger}},\ }\href
  {\doibase 10.1088/1361-6463/aa81a1} {\bibfield  {journal} {\bibinfo
  {journal} {Journal of Physics D: Applied Physics}\ }\textbf {\bibinfo
  {volume} {50}},\ \bibinfo {pages} {363001} (\bibinfo {year}
  {2017})}\BibitemShut {NoStop}%
\bibitem [{\citenamefont {Pirro}\ \emph {et~al.}(2021)\citenamefont {Pirro},
  \citenamefont {Vasyuchka}, \citenamefont {Serga},\ and\ \citenamefont
  {Hillebrands}}]{Pirro2021}%
  \BibitemOpen
  \bibfield  {author} {\bibinfo {author} {\bibfnamefont {P.}~\bibnamefont
  {Pirro}}, \bibinfo {author} {\bibfnamefont {V.~I.}\ \bibnamefont
  {Vasyuchka}}, \bibinfo {author} {\bibfnamefont {A.~A.}\ \bibnamefont
  {Serga}}, \ and\ \bibinfo {author} {\bibfnamefont {B.}~\bibnamefont
  {Hillebrands}},\ }\href {\doibase 10.1038/s41578-021-00332-w} {\bibfield
  {journal} {\bibinfo  {journal} {Nature Reviews Materials}\ }\textbf {\bibinfo
  {volume} {6}},\ \bibinfo {pages} {1114} (\bibinfo {year} {2021})}\BibitemShut
  {NoStop}%
\bibitem [{\citenamefont {Chumak}\ \emph {et~al.}(2022)\citenamefont {Chumak},
  \citenamefont {Kabos}, \citenamefont {Wu}, \citenamefont {Abert},
  \citenamefont {Adelmann}, \citenamefont {Adeyeye}, \citenamefont {Akerman},
  \citenamefont {Aliev}, \citenamefont {Anane}, \citenamefont {Awad},
  \citenamefont {Back}, \citenamefont {Barman}, \citenamefont {Bauer},
  \citenamefont {Becherer}, \citenamefont {Beginin}, \citenamefont
  {Bittencourt}, \citenamefont {Blanter}, \citenamefont {Bortolotti},
  \citenamefont {Boventer}, \citenamefont {Bozhko}, \citenamefont {Bunyaev},
  \citenamefont {Carmiggelt}, \citenamefont {Cheenikundil}, \citenamefont
  {Ciubotaru}, \citenamefont {Cotofana}, \citenamefont {Csaba}, \citenamefont
  {Dobrovolskiy}, \citenamefont {Dubs}, \citenamefont {Elyasi}, \citenamefont
  {Fripp}, \citenamefont {Fulara}, \citenamefont {Golovchanskiy}, \citenamefont
  {Gonzalez-Ballestero}, \citenamefont {Graczyk}, \citenamefont {Grundler},
  \citenamefont {Gruszecki}, \citenamefont {Gubbiotti}, \citenamefont
  {Guslienko}, \citenamefont {Haldar}, \citenamefont {Hamdioui}, \citenamefont
  {Hertel}, \citenamefont {Hillebrands}, \citenamefont {Hioki}, \citenamefont
  {Houshang}, \citenamefont {Hu}, \citenamefont {Huebl}, \citenamefont {Huth},
  \citenamefont {Iacocca}, \citenamefont {Jungfleisch}, \citenamefont
  {Kakazei}, \citenamefont {Khitun}, \citenamefont {Khymyn}, \citenamefont
  {Kikkawa}, \citenamefont {Klaui}, \citenamefont {Klein}, \citenamefont
  {Klos}, \citenamefont {Knauer}, \citenamefont {Koraltan}, \citenamefont
  {Kostylev}, \citenamefont {Krawczyk}, \citenamefont {Krivorotov},
  \citenamefont {Kruglyak}, \citenamefont {Lachance-Quirion}, \citenamefont
  {Ladak}, \citenamefont {Lebrun}, \citenamefont {Li}, \citenamefont {Lindner},
  \citenamefont {MacEdo}, \citenamefont {Mayr}, \citenamefont {Melkov},
  \citenamefont {Mieszczak}, \citenamefont {Nakamura}, \citenamefont {Nembach},
  \citenamefont {Nikitin}, \citenamefont {Nikitov}, \citenamefont {Novosad},
  \citenamefont {Otalora}, \citenamefont {Otani}, \citenamefont {Papp},
  \citenamefont {Pigeau}, \citenamefont {Pirro}, \citenamefont {Porod},
  \citenamefont {Porrati}, \citenamefont {Qin}, \citenamefont {Rana},
  \citenamefont {Reimann}, \citenamefont {Riente}, \citenamefont
  {Romero-Isart}, \citenamefont {Ross}, \citenamefont {Sadovnikov},
  \citenamefont {Safin}, \citenamefont {Saitoh}, \citenamefont {Schmidt},
  \citenamefont {Schultheiss}, \citenamefont {Schultheiss}, \citenamefont
  {Serga}, \citenamefont {Sharma}, \citenamefont {Shaw}, \citenamefont {Suess},
  \citenamefont {Surzhenko}, \citenamefont {Szulc}, \citenamefont {Taniguchi},
  \citenamefont {Urbanek}, \citenamefont {Usami}, \citenamefont {Ustinov},
  \citenamefont {{Van Der Sar}}, \citenamefont {{Van Dijken}}, \citenamefont
  {Vasyuchka}, \citenamefont {Verba}, \citenamefont {Kusminskiy}, \citenamefont
  {Wang}, \citenamefont {Weides}, \citenamefont {Weiler}, \citenamefont
  {Wintz}, \citenamefont {Wolski},\ and\ \citenamefont {Zhang}}]{Chumak2022}%
  \BibitemOpen
  \bibfield  {author} {\bibinfo {author} {\bibfnamefont {A.~V.}\ \bibnamefont
  {Chumak \emph {et~al.}}},\ }\href {\doibase 10.1109/TMAG.2022.3149664}
  {\bibfield  {journal} {\bibinfo  {journal} {IEEE Transactions on Magnetics}\
  }\textbf {\bibinfo {volume} {58}} (\bibinfo {year} {2022}),\
  10.1109/TMAG.2022.3149664}\BibitemShut {NoStop}%
\bibitem [{\citenamefont {Gomonay}\ \emph {et~al.}(2018)\citenamefont
  {Gomonay}, \citenamefont {Baltz}, \citenamefont {Brataas},\ and\
  \citenamefont {Tserkovnyak}}]{Gomonay2018}%
  \BibitemOpen
  \bibfield  {author} {\bibinfo {author} {\bibfnamefont {O.}~\bibnamefont
  {Gomonay}}, \bibinfo {author} {\bibfnamefont {V.}~\bibnamefont {Baltz}},
  \bibinfo {author} {\bibfnamefont {A.}~\bibnamefont {Brataas}}, \ and\
  \bibinfo {author} {\bibfnamefont {Y.}~\bibnamefont {Tserkovnyak}},\ }\href
  {\doibase 10.1038/s41567-018-0049-4} {\bibfield  {journal} {\bibinfo
  {journal} {Nature Physics}\ }\textbf {\bibinfo {volume} {14}},\ \bibinfo
  {pages} {213} (\bibinfo {year} {2018})}\BibitemShut {NoStop}%
\bibitem [{\citenamefont {Rezende}\ \emph {et~al.}(2019)\citenamefont
  {Rezende}, \citenamefont {Azevedo},\ and\ \citenamefont
  {Rodr{\'{i}}guez-Su{\'{a}}rez}}]{Rezende2019}%
  \BibitemOpen
  \bibfield  {author} {\bibinfo {author} {\bibfnamefont {S.~M.}\ \bibnamefont
  {Rezende}}, \bibinfo {author} {\bibfnamefont {A.}~\bibnamefont {Azevedo}}, \
  and\ \bibinfo {author} {\bibfnamefont {R.~L.}\ \bibnamefont
  {Rodr{\'{i}}guez-Su{\'{a}}rez}},\ }\href {\doibase 10.1063/1.5109132}
  {\bibfield  {journal} {\bibinfo  {journal} {Journal of Applied Physics}\
  }\textbf {\bibinfo {volume} {126}},\ \bibinfo {pages} {151101} (\bibinfo
  {year} {2019})}\BibitemShut {NoStop}%
\bibitem [{\citenamefont {{\v{S}}mejkal}\ \emph
  {et~al.}(2022{\natexlab{a}})\citenamefont {{\v{S}}mejkal}, \citenamefont
  {Sinova},\ and\ \citenamefont {Jungwirth}}]{Smejkal2021a}%
  \BibitemOpen
  \bibfield  {author} {\bibinfo {author} {\bibfnamefont {L.}~\bibnamefont
  {{\v{S}}mejkal}}, \bibinfo {author} {\bibfnamefont {J.}~\bibnamefont
  {Sinova}}, \ and\ \bibinfo {author} {\bibfnamefont {T.}~\bibnamefont
  {Jungwirth}},\ }\href {\doibase 10.1103/PhysRevX.12.031042} {\bibfield
  {journal} {\bibinfo  {journal} {Physical Review X}\ }\textbf {\bibinfo
  {volume} {12}},\ \bibinfo {pages} {031042} (\bibinfo {year}
  {2022}{\natexlab{a}})},\ \Eprint {http://arxiv.org/abs/2105.05820}
  {arXiv:2105.05820} \BibitemShut {NoStop}%
\bibitem [{\citenamefont {Smejkal}\ \emph {et~al.}(2022)\citenamefont
  {Smejkal}, \citenamefont {Sinova},\ and\ \citenamefont
  {Jungwirth}}]{Smejkal2022a}%
  \BibitemOpen
  \bibfield  {author} {\bibinfo {author} {\bibfnamefont {L.}~\bibnamefont
  {Smejkal}}, \bibinfo {author} {\bibfnamefont {J.}~\bibnamefont {Sinova}}, \
  and\ \bibinfo {author} {\bibfnamefont {T.}~\bibnamefont {Jungwirth}},\
  }\href@noop {} {\bibfield  {journal} {\bibinfo  {journal} {Phys. Rev. X}\ ,\
  \bibinfo {pages} {in press}} (\bibinfo {year} {2022})},\ \Eprint
  {http://arxiv.org/abs/2204.10844} {arXiv:2204.10844} \BibitemShut {NoStop}%
\bibitem [{\citenamefont {Halilov}\ \emph {et~al.}(1998)\citenamefont
  {Halilov}, \citenamefont {Eschrig}, \citenamefont {Perlov},\ and\
  \citenamefont {Oppeneer}}]{Halilov1998}%
  \BibitemOpen
  \bibfield  {author} {\bibinfo {author} {\bibfnamefont {S.~V.}\ \bibnamefont
  {Halilov}}, \bibinfo {author} {\bibfnamefont {H.}~\bibnamefont {Eschrig}},
  \bibinfo {author} {\bibfnamefont {A.~Y.}\ \bibnamefont {Perlov}}, \ and\
  \bibinfo {author} {\bibfnamefont {P.~M.}\ \bibnamefont {Oppeneer}},\ }\href
  {\doibase 10.1103/PhysRevB.58.293} {\bibfield  {journal} {\bibinfo  {journal}
  {Physical Review B}\ }\textbf {\bibinfo {volume} {58}},\ \bibinfo {pages}
  {293} (\bibinfo {year} {1998})}\BibitemShut {NoStop}%
\bibitem [{\citenamefont {Corticelli}\ \emph {et~al.}(2022)\citenamefont
  {Corticelli}, \citenamefont {Moessner},\ and\ \citenamefont
  {McClarty}}]{Corticelli2022}%
  \BibitemOpen
  \bibfield  {author} {\bibinfo {author} {\bibfnamefont {A.}~\bibnamefont
  {Corticelli}}, \bibinfo {author} {\bibfnamefont {R.}~\bibnamefont
  {Moessner}}, \ and\ \bibinfo {author} {\bibfnamefont {P.~A.}\ \bibnamefont
  {McClarty}},\ }\href {\doibase 10.1103/PhysRevB.105.064430} {\bibfield
  {journal} {\bibinfo  {journal} {Physical Review B}\ }\textbf {\bibinfo
  {volume} {105}},\ \bibinfo {pages} {064430} (\bibinfo {year} {2022})},\
  \Eprint {http://arxiv.org/abs/2103.05656} {arXiv:2103.05656} \BibitemShut
  {NoStop}%
\bibitem [{\citenamefont {Marmodoro}\ \emph
  {et~al.}(2022{\natexlab{a}})\citenamefont {Marmodoro}, \citenamefont
  {Mankovsky}, \citenamefont {Ebert}, \citenamefont {Min{\'{a}}r},\ and\
  \citenamefont {Ondřej{\v{s}}ipr}}]{Marmodoro2022}%
  \BibitemOpen
  \bibfield  {author} {\bibinfo {author} {\bibfnamefont {A.}~\bibnamefont
  {Marmodoro}}, \bibinfo {author} {\bibfnamefont {S.}~\bibnamefont
  {Mankovsky}}, \bibinfo {author} {\bibfnamefont {H.}~\bibnamefont {Ebert}},
  \bibinfo {author} {\bibfnamefont {J.}~\bibnamefont {Min{\'{a}}r}}, \ and\
  \bibinfo {author} {\bibfnamefont {O.~O.}\ \bibnamefont {Ondřej{\v{s}}ipr}},\
  }\href@noop {} {\  (\bibinfo {year} {2022}{\natexlab{a}})},\ \Eprint
  {http://arxiv.org/abs/2202.04525v1} {arXiv:2202.04525v1} \BibitemShut
  {NoStop}%
\bibitem [{\citenamefont {Brinkman}\ and\ \citenamefont
  {Elliott}(1966)}]{Brinkman1966}%
  \BibitemOpen
  \bibfield  {author} {\bibinfo {author} {\bibfnamefont {W.~F.}\ \bibnamefont
  {Brinkman}}\ and\ \bibinfo {author} {\bibfnamefont {R.~J.}\ \bibnamefont
  {Elliott}},\ }\href {\doibase 10.1098/rspa.1966.0211} {\bibfield  {journal}
  {\bibinfo  {journal} {Proceedings of the Royal Society A: Mathematical,
  Physical and Engineering Sciences}\ }\textbf {\bibinfo {volume} {294}},\
  \bibinfo {pages} {343} (\bibinfo {year} {1966})}\BibitemShut {NoStop}%
\bibitem [{\citenamefont {{\v{S}}mejkal}\ \emph {et~al.}(2020)\citenamefont
  {{\v{S}}mejkal}, \citenamefont {Gonz{\'{a}}lez-Hern{\'{a}}ndez},
  \citenamefont {Jungwirth},\ and\ \citenamefont {Sinova}}]{Smejkal2020}%
  \BibitemOpen
  \bibfield  {author} {\bibinfo {author} {\bibfnamefont {L.}~\bibnamefont
  {{\v{S}}mejkal}}, \bibinfo {author} {\bibfnamefont {R.}~\bibnamefont
  {Gonz{\'{a}}lez-Hern{\'{a}}ndez}}, \bibinfo {author} {\bibfnamefont
  {T.}~\bibnamefont {Jungwirth}}, \ and\ \bibinfo {author} {\bibfnamefont
  {J.}~\bibnamefont {Sinova}},\ }\href {\doibase 10.1126/sciadv.aaz8809}
  {\bibfield  {journal} {\bibinfo  {journal} {Science Advances}\ }\textbf
  {\bibinfo {volume} {6}},\ \bibinfo {pages} {eaaz8809} (\bibinfo {year}
  {2020})},\ \Eprint {http://arxiv.org/abs/1901.00445} {arXiv:1901.00445}
  \BibitemShut {NoStop}%
\bibitem [{\citenamefont {Feng}\ \emph {et~al.}(2022)\citenamefont {Feng},
  \citenamefont {Zhou}, \citenamefont {{\v{S}}mejkal}, \citenamefont {Wu},
  \citenamefont {Zhu}, \citenamefont {Guo}, \citenamefont
  {Gonz{\'{a}}lez-Hern{\'{a}}ndez}, \citenamefont {Wang}, \citenamefont {Yan},
  \citenamefont {Qin}, \citenamefont {Zhang}, \citenamefont {Wu}, \citenamefont
  {Chen}, \citenamefont {Meng}, \citenamefont {Liu}, \citenamefont {Xia},
  \citenamefont {Sinova}, \citenamefont {Jungwirth},\ and\ \citenamefont
  {Liu}}]{Feng2020a}%
  \BibitemOpen
  \bibfield  {author} {\bibinfo {author} {\bibfnamefont {Z.}~\bibnamefont
  {Feng}}, \bibinfo {author} {\bibfnamefont {X.}~\bibnamefont {Zhou}}, \bibinfo
  {author} {\bibfnamefont {L.}~\bibnamefont {{\v{S}}mejkal}}, \bibinfo {author}
  {\bibfnamefont {L.}~\bibnamefont {Wu}}, \bibinfo {author} {\bibfnamefont
  {Z.}~\bibnamefont {Zhu}}, \bibinfo {author} {\bibfnamefont {H.}~\bibnamefont
  {Guo}}, \bibinfo {author} {\bibfnamefont {R.}~\bibnamefont
  {Gonz{\'{a}}lez-Hern{\'{a}}ndez}}, \bibinfo {author} {\bibfnamefont
  {X.}~\bibnamefont {Wang}}, \bibinfo {author} {\bibfnamefont {H.}~\bibnamefont
  {Yan}}, \bibinfo {author} {\bibfnamefont {P.}~\bibnamefont {Qin}}, \bibinfo
  {author} {\bibfnamefont {X.}~\bibnamefont {Zhang}}, \bibinfo {author}
  {\bibfnamefont {H.}~\bibnamefont {Wu}}, \bibinfo {author} {\bibfnamefont
  {H.}~\bibnamefont {Chen}}, \bibinfo {author} {\bibfnamefont {Z.}~\bibnamefont
  {Meng}}, \bibinfo {author} {\bibfnamefont {L.}~\bibnamefont {Liu}}, \bibinfo
  {author} {\bibfnamefont {Z.}~\bibnamefont {Xia}}, \bibinfo {author}
  {\bibfnamefont {J.}~\bibnamefont {Sinova}}, \bibinfo {author} {\bibfnamefont
  {T.}~\bibnamefont {Jungwirth}}, \ and\ \bibinfo {author} {\bibfnamefont
  {Z.}~\bibnamefont {Liu}},\ }\href {\doibase 10.1038/s41928-022-00866-z}
  {\bibfield  {journal} {\bibinfo  {journal} {Nature Electronics}\ ,\ \bibinfo
  {pages} {published 7 Nov}} (\bibinfo {year} {2022})},\ \Eprint
  {http://arxiv.org/abs/2002.08712} {arXiv:2002.08712} \BibitemShut {NoStop}%
\bibitem [{\citenamefont {{\v{S}}mejkal}\ \emph
  {et~al.}(2022{\natexlab{b}})\citenamefont {{\v{S}}mejkal}, \citenamefont
  {MacDonald}, \citenamefont {Sinova}, \citenamefont {Nakatsuji},\ and\
  \citenamefont {Jungwirth}}]{Smejkal2021b}%
  \BibitemOpen
  \bibfield  {author} {\bibinfo {author} {\bibfnamefont {L.}~\bibnamefont
  {{\v{S}}mejkal}}, \bibinfo {author} {\bibfnamefont {A.~H.}\ \bibnamefont
  {MacDonald}}, \bibinfo {author} {\bibfnamefont {J.}~\bibnamefont {Sinova}},
  \bibinfo {author} {\bibfnamefont {S.}~\bibnamefont {Nakatsuji}}, \ and\
  \bibinfo {author} {\bibfnamefont {T.}~\bibnamefont {Jungwirth}},\ }\href
  {\doibase 10.1038/s41578-022-00430-3} {\bibfield  {journal} {\bibinfo
  {journal} {Nature Reviews Materials}\ }\textbf {\bibinfo {volume} {7}},\
  \bibinfo {pages} {482} (\bibinfo {year} {2022}{\natexlab{b}})},\ \Eprint
  {http://arxiv.org/abs/2107.03321} {arXiv:2107.03321} \BibitemShut {NoStop}%
\bibitem [{\citenamefont {Gonz{\'{a}}lez-Hern{\'{a}}ndez}\ \emph
  {et~al.}(2021)\citenamefont {Gonz{\'{a}}lez-Hern{\'{a}}ndez}, \citenamefont
  {{\v{S}}mejkal}, \citenamefont {V{\'{y}}born{\'{y}}}, \citenamefont {Yahagi},
  \citenamefont {Sinova}, \citenamefont {Jungwirth},\ and\ \citenamefont
  {{\v{Z}}elezn{\'{y}}}}]{Gonzalez-Hernandez2021}%
  \BibitemOpen
  \bibfield  {author} {\bibinfo {author} {\bibfnamefont {R.}~\bibnamefont
  {Gonz{\'{a}}lez-Hern{\'{a}}ndez}}, \bibinfo {author} {\bibfnamefont
  {L.}~\bibnamefont {{\v{S}}mejkal}}, \bibinfo {author} {\bibfnamefont
  {K.}~\bibnamefont {V{\'{y}}born{\'{y}}}}, \bibinfo {author} {\bibfnamefont
  {Y.}~\bibnamefont {Yahagi}}, \bibinfo {author} {\bibfnamefont
  {J.}~\bibnamefont {Sinova}}, \bibinfo {author} {\bibfnamefont
  {T.}~\bibnamefont {Jungwirth}}, \ and\ \bibinfo {author} {\bibfnamefont
  {J.}~\bibnamefont {{\v{Z}}elezn{\'{y}}}},\ }\href {\doibase
  10.1103/PhysRevLett.126.127701} {\bibfield  {journal} {\bibinfo  {journal}
  {Physical Review Letters}\ }\textbf {\bibinfo {volume} {126}},\ \bibinfo
  {pages} {127701} (\bibinfo {year} {2021})},\ \Eprint
  {http://arxiv.org/abs/2002.07073} {arXiv:2002.07073} \BibitemShut {NoStop}%
\bibitem [{\citenamefont {{\v{S}}mejkal}\ \emph
  {et~al.}(2022{\natexlab{c}})\citenamefont {{\v{S}}mejkal}, \citenamefont
  {Hellenes}, \citenamefont {Gonz{\'{a}}lez-Hern{\'{a}}ndez}, \citenamefont
  {Sinova},\ and\ \citenamefont {Jungwirth}}]{Smejkal2022}%
  \BibitemOpen
  \bibfield  {author} {\bibinfo {author} {\bibfnamefont {L.}~\bibnamefont
  {{\v{S}}mejkal}}, \bibinfo {author} {\bibfnamefont {A.~B.}\ \bibnamefont
  {Hellenes}}, \bibinfo {author} {\bibfnamefont {R.}~\bibnamefont
  {Gonz{\'{a}}lez-Hern{\'{a}}ndez}}, \bibinfo {author} {\bibfnamefont
  {J.}~\bibnamefont {Sinova}}, \ and\ \bibinfo {author} {\bibfnamefont
  {T.}~\bibnamefont {Jungwirth}},\ }\href {\doibase 10.1103/PhysRevX.12.011028}
  {\bibfield  {journal} {\bibinfo  {journal} {Physical Review X}\ }\textbf
  {\bibinfo {volume} {12}},\ \bibinfo {pages} {011028} (\bibinfo {year}
  {2022}{\natexlab{c}})},\ \Eprint {http://arxiv.org/abs/2103.12664}
  {arXiv:2103.12664} \BibitemShut {NoStop}%
\bibitem [{\citenamefont {Shao}\ \emph {et~al.}(2021)\citenamefont {Shao},
  \citenamefont {Zhang}, \citenamefont {Li}, \citenamefont {Eom},\ and\
  \citenamefont {Tsymbal}}]{Shao2021}%
  \BibitemOpen
  \bibfield  {author} {\bibinfo {author} {\bibfnamefont {D.-F.}\ \bibnamefont
  {Shao}}, \bibinfo {author} {\bibfnamefont {S.-H.}\ \bibnamefont {Zhang}},
  \bibinfo {author} {\bibfnamefont {M.}~\bibnamefont {Li}}, \bibinfo {author}
  {\bibfnamefont {C.-B.}\ \bibnamefont {Eom}}, \ and\ \bibinfo {author}
  {\bibfnamefont {E.~Y.}\ \bibnamefont {Tsymbal}},\ }\href {\doibase
  10.1038/s41467-021-26915-3} {\bibfield  {journal} {\bibinfo  {journal}
  {Nature Communications}\ }\textbf {\bibinfo {volume} {12}},\ \bibinfo {pages}
  {7061} (\bibinfo {year} {2021})},\ \Eprint {http://arxiv.org/abs/2103.09219}
  {arXiv:2103.09219} \BibitemShut {NoStop}%
\bibitem [{\citenamefont {Ahn}\ \emph {et~al.}(2019)\citenamefont {Ahn},
  \citenamefont {Hariki}, \citenamefont {Lee},\ and\ \citenamefont
  {Kune{\v{s}}}}]{Ahn2019}%
  \BibitemOpen
  \bibfield  {author} {\bibinfo {author} {\bibfnamefont {K.-H.}\ \bibnamefont
  {Ahn}}, \bibinfo {author} {\bibfnamefont {A.}~\bibnamefont {Hariki}},
  \bibinfo {author} {\bibfnamefont {K.-W.}\ \bibnamefont {Lee}}, \ and\
  \bibinfo {author} {\bibfnamefont {J.}~\bibnamefont {Kune{\v{s}}}},\ }\href
  {\doibase 10.1103/PhysRevB.99.184432} {\bibfield  {journal} {\bibinfo
  {journal} {Physical Review B}\ }\textbf {\bibinfo {volume} {99}},\ \bibinfo
  {pages} {184432} (\bibinfo {year} {2019})},\ \Eprint
  {http://arxiv.org/abs/1902.04436} {arXiv:1902.04436} \BibitemShut {NoStop}%
\bibitem [{\citenamefont {N{\'{e}}el}(1953)}]{Neel1953}%
  \BibitemOpen
  \bibfield  {author} {\bibinfo {author} {\bibfnamefont {L.}~\bibnamefont
  {N{\'{e}}el}},\ }\href {\doibase 10.1103/RevModPhys.25.58} {\bibfield
  {journal} {\bibinfo  {journal} {Reviews of Modern Physics}\ }\textbf
  {\bibinfo {volume} {25}},\ \bibinfo {pages} {58} (\bibinfo {year}
  {1953})}\BibitemShut {NoStop}%
\bibitem [{\citenamefont {Bose}\ \emph {et~al.}(2022)\citenamefont {Bose},
  \citenamefont {Schreiber}, \citenamefont {Jain}, \citenamefont {Shao},
  \citenamefont {Nair}, \citenamefont {Sun}, \citenamefont {Zhang},
  \citenamefont {Muller}, \citenamefont {Tsymbal}, \citenamefont {Schlom},\
  and\ \citenamefont {Ralph}}]{Bose2022}%
  \BibitemOpen
  \bibfield  {author} {\bibinfo {author} {\bibfnamefont {A.}~\bibnamefont
  {Bose}}, \bibinfo {author} {\bibfnamefont {N.~J.}\ \bibnamefont {Schreiber}},
  \bibinfo {author} {\bibfnamefont {R.}~\bibnamefont {Jain}}, \bibinfo {author}
  {\bibfnamefont {D.-f.}\ \bibnamefont {Shao}}, \bibinfo {author}
  {\bibfnamefont {H.~P.}\ \bibnamefont {Nair}}, \bibinfo {author}
  {\bibfnamefont {J.}~\bibnamefont {Sun}}, \bibinfo {author} {\bibfnamefont
  {X.~S.}\ \bibnamefont {Zhang}}, \bibinfo {author} {\bibfnamefont {D.~A.}\
  \bibnamefont {Muller}}, \bibinfo {author} {\bibfnamefont {E.~Y.}\
  \bibnamefont {Tsymbal}}, \bibinfo {author} {\bibfnamefont {D.~G.}\
  \bibnamefont {Schlom}}, \ and\ \bibinfo {author} {\bibfnamefont {D.~C.}\
  \bibnamefont {Ralph}},\ }\href {\doibase 10.1038/s41928-022-00758-2}
  {\bibfield  {journal} {\bibinfo  {journal} {Nature Electronics}\ }\textbf
  {\bibinfo {volume} {5}},\ \bibinfo {pages} {263} (\bibinfo {year} {2022})},\
  \Eprint {http://arxiv.org/abs/2108.09150} {arXiv:2108.09150} \BibitemShut
  {NoStop}%
\bibitem [{\citenamefont {Bai}\ \emph {et~al.}(2022)\citenamefont {Bai},
  \citenamefont {Han}, \citenamefont {Feng}, \citenamefont {Zhou},
  \citenamefont {Su}, \citenamefont {Wang}, \citenamefont {Liao}, \citenamefont
  {Zhu}, \citenamefont {Chen}, \citenamefont {Pan}, \citenamefont {Fan},\ and\
  \citenamefont {Song}}]{Bai2021}%
  \BibitemOpen
  \bibfield  {author} {\bibinfo {author} {\bibfnamefont {H.}~\bibnamefont
  {Bai}}, \bibinfo {author} {\bibfnamefont {L.}~\bibnamefont {Han}}, \bibinfo
  {author} {\bibfnamefont {X.~Y.}\ \bibnamefont {Feng}}, \bibinfo {author}
  {\bibfnamefont {Y.~J.}\ \bibnamefont {Zhou}}, \bibinfo {author}
  {\bibfnamefont {R.~X.}\ \bibnamefont {Su}}, \bibinfo {author} {\bibfnamefont
  {Q.}~\bibnamefont {Wang}}, \bibinfo {author} {\bibfnamefont {L.~Y.}\
  \bibnamefont {Liao}}, \bibinfo {author} {\bibfnamefont {W.~X.}\ \bibnamefont
  {Zhu}}, \bibinfo {author} {\bibfnamefont {X.~Z.}\ \bibnamefont {Chen}},
  \bibinfo {author} {\bibfnamefont {F.}~\bibnamefont {Pan}}, \bibinfo {author}
  {\bibfnamefont {X.~L.}\ \bibnamefont {Fan}}, \ and\ \bibinfo {author}
  {\bibfnamefont {C.}~\bibnamefont {Song}},\ }\href {\doibase
  10.1103/PhysRevLett.128.197202} {\bibfield  {journal} {\bibinfo  {journal}
  {Physical Review Letters}\ }\textbf {\bibinfo {volume} {128}},\ \bibinfo
  {pages} {197202} (\bibinfo {year} {2022})},\ \Eprint
  {http://arxiv.org/abs/2109.05933} {arXiv:2109.05933} \BibitemShut {NoStop}%
\bibitem [{\citenamefont {Karube}\ \emph {et~al.}(2022)\citenamefont {Karube},
  \citenamefont {Tanaka}, \citenamefont {Sugawara}, \citenamefont {Kadoguchi},
  \citenamefont {Kohda},\ and\ \citenamefont {Nitta}}]{Karube2022}%
  \BibitemOpen
  \bibfield  {author} {\bibinfo {author} {\bibfnamefont {S.}~\bibnamefont
  {Karube}}, \bibinfo {author} {\bibfnamefont {T.}~\bibnamefont {Tanaka}},
  \bibinfo {author} {\bibfnamefont {D.}~\bibnamefont {Sugawara}}, \bibinfo
  {author} {\bibfnamefont {N.}~\bibnamefont {Kadoguchi}}, \bibinfo {author}
  {\bibfnamefont {M.}~\bibnamefont {Kohda}}, \ and\ \bibinfo {author}
  {\bibfnamefont {J.}~\bibnamefont {Nitta}},\ }\href {\doibase
  10.1103/PhysRevLett.129.137201} {\bibfield  {journal} {\bibinfo  {journal}
  {Physical Review Letters}\ }\textbf {\bibinfo {volume} {129}},\ \bibinfo
  {pages} {137201} (\bibinfo {year} {2022})},\ \Eprint
  {http://arxiv.org/abs/2111.07487} {arXiv:2111.07487} \BibitemShut {NoStop}%
\bibitem [{\citenamefont {Samanta}\ \emph {et~al.}(2020)\citenamefont
  {Samanta}, \citenamefont {Le{\v{z}}ai{\'{c}}}, \citenamefont {Merte},
  \citenamefont {Freimuth}, \citenamefont {Bl{\"{u}}gel},\ and\ \citenamefont
  {Mokrousov}}]{Samanta2020}%
  \BibitemOpen
  \bibfield  {author} {\bibinfo {author} {\bibfnamefont {K.}~\bibnamefont
  {Samanta}}, \bibinfo {author} {\bibfnamefont {M.}~\bibnamefont
  {Le{\v{z}}ai{\'{c}}}}, \bibinfo {author} {\bibfnamefont {M.}~\bibnamefont
  {Merte}}, \bibinfo {author} {\bibfnamefont {F.}~\bibnamefont {Freimuth}},
  \bibinfo {author} {\bibfnamefont {S.}~\bibnamefont {Bl{\"{u}}gel}}, \ and\
  \bibinfo {author} {\bibfnamefont {Y.}~\bibnamefont {Mokrousov}},\ }\href
  {\doibase 10.1063/5.0005017} {\bibfield  {journal} {\bibinfo  {journal}
  {Journal of Applied Physics}\ }\textbf {\bibinfo {volume} {127}},\ \bibinfo
  {pages} {213904} (\bibinfo {year} {2020})}\BibitemShut {NoStop}%
\bibitem [{\citenamefont {Naka}\ \emph {et~al.}(2020)\citenamefont {Naka},
  \citenamefont {Hayami}, \citenamefont {Kusunose}, \citenamefont {Yanagi},
  \citenamefont {Motome},\ and\ \citenamefont {Seo}}]{Naka2020}%
  \BibitemOpen
  \bibfield  {author} {\bibinfo {author} {\bibfnamefont {M.}~\bibnamefont
  {Naka}}, \bibinfo {author} {\bibfnamefont {S.}~\bibnamefont {Hayami}},
  \bibinfo {author} {\bibfnamefont {H.}~\bibnamefont {Kusunose}}, \bibinfo
  {author} {\bibfnamefont {Y.}~\bibnamefont {Yanagi}}, \bibinfo {author}
  {\bibfnamefont {Y.}~\bibnamefont {Motome}}, \ and\ \bibinfo {author}
  {\bibfnamefont {H.}~\bibnamefont {Seo}},\ }\href {\doibase
  10.1103/PhysRevB.102.075112} {\bibfield  {journal} {\bibinfo  {journal}
  {Physical Review B}\ }\textbf {\bibinfo {volume} {102}},\ \bibinfo {pages}
  {075112} (\bibinfo {year} {2020})}\BibitemShut {NoStop}%
\bibitem [{\citenamefont {Reichlova}\ \emph {et~al.}(2020)\citenamefont
  {Reichlova}, \citenamefont {{Lopes Seeger}}, \citenamefont
  {Gonz{\'{a}}lez-Hern{\'{a}}ndez}, \citenamefont {Kounta}, \citenamefont
  {Schlitz}, \citenamefont {Kriegner}, \citenamefont {Ritzinger}, \citenamefont
  {Lammel}, \citenamefont {Leivisk{\"{a}}}, \citenamefont {Petři{\v{c}}ek},
  \citenamefont {Dole{\v{z}}al}, \citenamefont {Schmoranzerova}, \citenamefont
  {Bad}, \citenamefont {Thomas}, \citenamefont {Baltz}, \citenamefont {Michez},
  \citenamefont {Sinova}, \citenamefont {{B Goennenwein}}, \citenamefont
  {Jungwirth},\ and\ \citenamefont {Smejkal}}]{Reichlova2020}%
  \BibitemOpen
  \bibfield  {author} {\bibinfo {author} {\bibfnamefont {H.}~\bibnamefont
  {Reichlova}}, \bibinfo {author} {\bibfnamefont {R.}~\bibnamefont {{Lopes
  Seeger}}}, \bibinfo {author} {\bibfnamefont {R.}~\bibnamefont
  {Gonz{\'{a}}lez-Hern{\'{a}}ndez}}, \bibinfo {author} {\bibfnamefont
  {I.}~\bibnamefont {Kounta}}, \bibinfo {author} {\bibfnamefont
  {R.}~\bibnamefont {Schlitz}}, \bibinfo {author} {\bibfnamefont
  {D.}~\bibnamefont {Kriegner}}, \bibinfo {author} {\bibfnamefont
  {P.}~\bibnamefont {Ritzinger}}, \bibinfo {author} {\bibfnamefont
  {M.}~\bibnamefont {Lammel}}, \bibinfo {author} {\bibfnamefont
  {M.}~\bibnamefont {Leivisk{\"{a}}}}, \bibinfo {author} {\bibfnamefont
  {V.}~\bibnamefont {Petři{\v{c}}ek}}, \bibinfo {author} {\bibfnamefont
  {P.}~\bibnamefont {Dole{\v{z}}al}}, \bibinfo {author} {\bibfnamefont
  {E.}~\bibnamefont {Schmoranzerova}}, \bibinfo {author} {\bibfnamefont
  {A.}~\bibnamefont {Bad}}, \bibinfo {author} {\bibfnamefont {A.}~\bibnamefont
  {Thomas}}, \bibinfo {author} {\bibfnamefont {V.}~\bibnamefont {Baltz}},
  \bibinfo {author} {\bibfnamefont {L.}~\bibnamefont {Michez}}, \bibinfo
  {author} {\bibfnamefont {J.}~\bibnamefont {Sinova}}, \bibinfo {author}
  {\bibfnamefont {S.~T.}\ \bibnamefont {{B Goennenwein}}}, \bibinfo {author}
  {\bibfnamefont {T.}~\bibnamefont {Jungwirth}}, \ and\ \bibinfo {author}
  {\bibfnamefont {L.}~\bibnamefont {Smejkal}},\ }\href
  {https://arxiv.org/pdf/2012.15651.pdf} {\emph {\bibinfo {title} {{Macroscopic
  time reversal symmetry breaking arising from antiferromagnetic Zeeman
  effect}}}},\ \bibinfo {type} {Tech. Rep.}\ (\bibinfo {year} {2020})\ \Eprint
  {http://arxiv.org/abs/2012.15651v1} {arXiv:2012.15651v1} \BibitemShut
  {NoStop}%
\bibitem [{\citenamefont {Mazin}\ \emph {et~al.}(2021)\citenamefont {Mazin},
  \citenamefont {Koepernik}, \citenamefont {Johannes}, \citenamefont
  {Gonz{\'{a}}lez-Hern{\'{a}}ndez},\ and\ \citenamefont
  {{\v{S}}mejkal}}]{Mazin2021}%
  \BibitemOpen
  \bibfield  {author} {\bibinfo {author} {\bibfnamefont {I.~I.}\ \bibnamefont
  {Mazin}}, \bibinfo {author} {\bibfnamefont {K.}~\bibnamefont {Koepernik}},
  \bibinfo {author} {\bibfnamefont {M.~D.}\ \bibnamefont {Johannes}}, \bibinfo
  {author} {\bibfnamefont {R.}~\bibnamefont {Gonz{\'{a}}lez-Hern{\'{a}}ndez}},
  \ and\ \bibinfo {author} {\bibfnamefont {L.}~\bibnamefont {{\v{S}}mejkal}},\
  }\href {http://arxiv.org/abs/2105.06356} {\bibfield  {journal} {\bibinfo
  {journal} {PNAS}\ }\textbf {\bibinfo {volume} {118}},\ \bibinfo {pages}
  {e2108924118} (\bibinfo {year} {2021})},\ \Eprint
  {http://arxiv.org/abs/2105.06356} {arXiv:2105.06356} \BibitemShut {NoStop}%
\bibitem [{\citenamefont {Betancourt}\ \emph {et~al.}(2021)\citenamefont
  {Betancourt}, \citenamefont {Zub{\'{a}}{\v{c}}}, \citenamefont
  {Gonzalez-Hernandez}, \citenamefont {Geishendorf}, \citenamefont
  {{\v{S}}ob{\'{a}}ň}, \citenamefont {Springholz}, \citenamefont
  {V{\'{y}}born{\'{y}}}, \citenamefont {Olejn{\'{i}}k}, \citenamefont
  {{\v{S}}mejkal}, \citenamefont {Sinova}, \citenamefont {Jungwirth},
  \citenamefont {Goennenwein}, \citenamefont {Thomas}, \citenamefont
  {Reichlov{\'{a}}}, \citenamefont {{\v{Z}}elezn{\'{y}}},\ and\ \citenamefont
  {Kriegner}}]{Betancourt2021}%
  \BibitemOpen
  \bibfield  {author} {\bibinfo {author} {\bibfnamefont {R.~D.~G.}\
  \bibnamefont {Betancourt}}, \bibinfo {author} {\bibfnamefont
  {J.}~\bibnamefont {Zub{\'{a}}{\v{c}}}}, \bibinfo {author} {\bibfnamefont
  {R.~J.}\ \bibnamefont {Gonzalez-Hernandez}}, \bibinfo {author} {\bibfnamefont
  {K.}~\bibnamefont {Geishendorf}}, \bibinfo {author} {\bibfnamefont
  {Z.}~\bibnamefont {{\v{S}}ob{\'{a}}ň}}, \bibinfo {author} {\bibfnamefont
  {G.}~\bibnamefont {Springholz}}, \bibinfo {author} {\bibfnamefont
  {K.}~\bibnamefont {V{\'{y}}born{\'{y}}}}, \bibinfo {author} {\bibfnamefont
  {K.}~\bibnamefont {Olejn{\'{i}}k}}, \bibinfo {author} {\bibfnamefont
  {L.}~\bibnamefont {{\v{S}}mejkal}}, \bibinfo {author} {\bibfnamefont
  {J.}~\bibnamefont {Sinova}}, \bibinfo {author} {\bibfnamefont
  {T.}~\bibnamefont {Jungwirth}}, \bibinfo {author} {\bibfnamefont {S.~T.~B.}\
  \bibnamefont {Goennenwein}}, \bibinfo {author} {\bibfnamefont
  {A.}~\bibnamefont {Thomas}}, \bibinfo {author} {\bibfnamefont
  {H.}~\bibnamefont {Reichlov{\'{a}}}}, \bibinfo {author} {\bibfnamefont
  {J.}~\bibnamefont {{\v{Z}}elezn{\'{y}}}}, \ and\ \bibinfo {author}
  {\bibfnamefont {D.}~\bibnamefont {Kriegner}},\ }\href
  {https://arxiv.org/abs/2112.06805v1} {\  (\bibinfo {year} {2021})},\ \Eprint
  {http://arxiv.org/abs/2112.06805} {arXiv:2112.06805} \BibitemShut {NoStop}%
\bibitem [{\citenamefont {Naka}\ \emph {et~al.}(2022)\citenamefont {Naka},
  \citenamefont {Motome},\ and\ \citenamefont {Seo}}]{Naka2022}%
  \BibitemOpen
  \bibfield  {author} {\bibinfo {author} {\bibfnamefont {M.}~\bibnamefont
  {Naka}}, \bibinfo {author} {\bibfnamefont {Y.}~\bibnamefont {Motome}}, \ and\
  \bibinfo {author} {\bibfnamefont {H.}~\bibnamefont {Seo}},\ }\href
  {http://arxiv.org/abs/2208.11823} {\ (\bibinfo {year}
  {2022})},\ \Eprint {http://arxiv.org/abs/2208.11823} {arXiv:2208.11823}
  \BibitemShut {NoStop}%
\bibitem [{\citenamefont {Costa}\ \emph {et~al.}(2003)\citenamefont {Costa},
  \citenamefont {Muniz},\ and\ \citenamefont {Mills}}]{Costa2003}%
  \BibitemOpen
  \bibfield  {author} {\bibinfo {author} {\bibfnamefont {A.~T.}\ \bibnamefont
  {Costa}}, \bibinfo {author} {\bibfnamefont {R.~B.}\ \bibnamefont {Muniz}}, \
  and\ \bibinfo {author} {\bibfnamefont {D.~L.}\ \bibnamefont {Mills}},\ }\href
  {\doibase 10.1103/PhysRevB.68.224435} {\bibfield  {journal} {\bibinfo
  {journal} {Physical Review B}\ }\textbf {\bibinfo {volume} {68}},\ \bibinfo
  {pages} {224435} (\bibinfo {year} {2003})}\BibitemShut {NoStop}%
\bibitem [{\citenamefont {Marmodoro}\ \emph
  {et~al.}(2022{\natexlab{b}})\citenamefont {Marmodoro}, \citenamefont
  {Mankovsky}, \citenamefont {Ebert}, \citenamefont {Min{\'{a}}r},\ and\
  \citenamefont {{\v{S}}ipr}}]{Marmodoro2022a}%
  \BibitemOpen
  \bibfield  {author} {\bibinfo {author} {\bibfnamefont {A.}~\bibnamefont
  {Marmodoro}}, \bibinfo {author} {\bibfnamefont {S.}~\bibnamefont
  {Mankovsky}}, \bibinfo {author} {\bibfnamefont {H.}~\bibnamefont {Ebert}},
  \bibinfo {author} {\bibfnamefont {J.}~\bibnamefont {Min{\'{a}}r}}, \ and\
  \bibinfo {author} {\bibfnamefont {O.}~\bibnamefont {{\v{S}}ipr}},\ }\href
  {\doibase 10.1103/PhysRevB.105.174411} {\bibfield  {journal} {\bibinfo
  {journal} {Physical Review B}\ }\textbf {\bibinfo {volume} {105}},\ \bibinfo
  {pages} {174411} (\bibinfo {year} {2022}{\natexlab{b}})}\BibitemShut
  {NoStop}%
\bibitem [{\citenamefont {Yosida}(1996)}]{Yosida1996}%
  \BibitemOpen
  \bibfield  {author} {\bibinfo {author} {\bibfnamefont {K.}~\bibnamefont
  {Yosida}},\ }\href@noop {} {\emph {\bibinfo {title} {{Theory of
  Magnetism}}}}\ (\bibinfo  {publisher} {Springer Berlin, Heidelberg},\
  \bibinfo {year} {1996})\BibitemShut {NoStop}%
\bibitem [{\citenamefont {Kakehashi}(2013)}]{Kakehashi2013}%
  \BibitemOpen
  \bibfield  {author} {\bibinfo {author} {\bibfnamefont {Y.}~\bibnamefont
  {Kakehashi}},\ }\href {\doibase 10.1007/978-3-642-33401-6} {\emph {\bibinfo
  {title} {{Modern Theory of Magnetism in Metals and Alloys}}}},\ \bibinfo
  {series} {Springer Series in Solid-State Sciences}, Vol.\ \bibinfo {volume}
  {175}\ (\bibinfo  {publisher} {Springer Berlin Heidelberg},\ \bibinfo
  {address} {Berlin, Heidelberg},\ \bibinfo {year} {2013})\BibitemShut
  {NoStop}%
\bibitem [{\citenamefont {Haines}\ \emph {et~al.}(1997)\citenamefont {Haines},
  \citenamefont {L{\'{e}}ger}, \citenamefont {Schulte},\ and\ \citenamefont
  {Hull}}]{Haines1997}%
  \BibitemOpen
  \bibfield  {author} {\bibinfo {author} {\bibfnamefont {J.}~\bibnamefont
  {Haines}}, \bibinfo {author} {\bibfnamefont {J.~M.}\ \bibnamefont
  {L{\'{e}}ger}}, \bibinfo {author} {\bibfnamefont {O.}~\bibnamefont
  {Schulte}}, \ and\ \bibinfo {author} {\bibfnamefont {S.}~\bibnamefont
  {Hull}},\ }\href {\doibase 10.1107/S0108768197008094} {\bibfield  {journal}
  {\bibinfo  {journal} {Acta Crystallographica Section B Structural Science}\
  }\textbf {\bibinfo {volume} {53}},\ \bibinfo {pages} {880} (\bibinfo {year}
  {1997})}\BibitemShut {NoStop}%
\bibitem [{\citenamefont {Ebert}\ \emph {et~al.}(2011)\citenamefont {Ebert},
  \citenamefont {K{\"{o}}dderitzsch},\ and\ \citenamefont
  {Min{\'{a}}r}}]{Ebert2011a}%
  \BibitemOpen
  \bibfield  {author} {\bibinfo {author} {\bibfnamefont {H.}~\bibnamefont
  {Ebert}}, \bibinfo {author} {\bibfnamefont {D.}~\bibnamefont
  {K{\"{o}}dderitzsch}}, \ and\ \bibinfo {author} {\bibfnamefont
  {J.}~\bibnamefont {Min{\'{a}}r}},\ }\href {\doibase
  10.1088/0034-4885/74/9/096501} {\bibfield  {journal} {\bibinfo  {journal}
  {Reports on Progress in Physics}\ }\textbf {\bibinfo {volume} {74}},\
  \bibinfo {pages} {096501} (\bibinfo {year} {2011})}\BibitemShut {NoStop}%
\bibitem [{\citenamefont {Vosko}\ \emph {et~al.}(1980)\citenamefont {Vosko},
  \citenamefont {Wilk},\ and\ \citenamefont {Nusair}}]{Vosko1980}%
  \BibitemOpen
  \bibfield  {author} {\bibinfo {author} {\bibfnamefont {S.~H.}\ \bibnamefont
  {Vosko}}, \bibinfo {author} {\bibfnamefont {L.}~\bibnamefont {Wilk}}, \ and\
  \bibinfo {author} {\bibfnamefont {M.}~\bibnamefont {Nusair}},\ }\href
  {\doibase 10.1139/p80-159} {\bibfield  {journal} {\bibinfo  {journal}
  {Canadian Journal of Physics}\ }\textbf {\bibinfo {volume} {58}},\ \bibinfo
  {pages} {1200} (\bibinfo {year} {1980})}\BibitemShut {NoStop}%
\bibitem [{\citenamefont {Dudarev}\ \emph {et~al.}(1998)\citenamefont
  {Dudarev}, \citenamefont {Botton}, \citenamefont {Savrasov}, \citenamefont
  {Humphreys},\ and\ \citenamefont {Sutton}}]{Dudarev1998}%
  \BibitemOpen
  \bibfield  {author} {\bibinfo {author} {\bibfnamefont {S.~L.}\ \bibnamefont
  {Dudarev}}, \bibinfo {author} {\bibfnamefont {G.~A.}\ \bibnamefont {Botton}},
  \bibinfo {author} {\bibfnamefont {S.~Y.}\ \bibnamefont {Savrasov}}, \bibinfo
  {author} {\bibfnamefont {C.~J.}\ \bibnamefont {Humphreys}}, \ and\ \bibinfo
  {author} {\bibfnamefont {A.~P.}\ \bibnamefont {Sutton}},\ }\href {\doibase
  10.1103/PhysRevB.57.1505} {\bibfield  {journal} {\bibinfo  {journal}
  {Physical Review B}\ }\textbf {\bibinfo {volume} {57}},\ \bibinfo {pages}
  {1505} (\bibinfo {year} {1998})}\BibitemShut {NoStop}%
\bibitem [{\citenamefont {Strange}\ \emph {et~al.}(1984)\citenamefont
  {Strange}, \citenamefont {Staunton},\ and\ \citenamefont
  {Gyorffy}}]{Strange1984}%
  \BibitemOpen
  \bibfield  {author} {\bibinfo {author} {\bibfnamefont {P.}~\bibnamefont
  {Strange}}, \bibinfo {author} {\bibfnamefont {J.}~\bibnamefont {Staunton}}, \
  and\ \bibinfo {author} {\bibfnamefont {B.~L.}\ \bibnamefont {Gyorffy}},\
  }\href {\doibase 10.1088/0022-3719/17/19/011} {\bibfield  {journal} {\bibinfo
   {journal} {Journal of Physics C: Solid State Physics}\ }\textbf {\bibinfo
  {volume} {17}},\ \bibinfo {pages} {3355} (\bibinfo {year}
  {1984})}\BibitemShut {NoStop}%
\bibitem [{\citenamefont {Mankovsky}\ and\ \citenamefont
  {Ebert}(2017)}]{Mankovsky2017}%
  \BibitemOpen
  \bibfield  {author} {\bibinfo {author} {\bibfnamefont {S.}~\bibnamefont
  {Mankovsky}}\ and\ \bibinfo {author} {\bibfnamefont {H.}~\bibnamefont
  {Ebert}},\ }\href {\doibase 10.1103/PhysRevB.96.104416} {\bibfield  {journal}
  {\bibinfo  {journal} {Physical Review B}\ }\textbf {\bibinfo {volume} {96}},\
  \bibinfo {pages} {104416} (\bibinfo {year} {2017})},\ \Eprint
  {http://arxiv.org/abs/1706.04165} {arXiv:1706.04165} \BibitemShut {NoStop}%
\bibitem [{\citenamefont {Berlijn}\ \emph {et~al.}(2017)\citenamefont
  {Berlijn}, \citenamefont {Snijders}, \citenamefont {Delaire}, \citenamefont
  {Zhou}, \citenamefont {Maier}, \citenamefont {Cao}, \citenamefont {Chi},
  \citenamefont {Matsuda}, \citenamefont {Wang}, \citenamefont {Koehler},
  \citenamefont {Kent},\ and\ \citenamefont {Weitering}}]{Berlijn2017a}%
  \BibitemOpen
  \bibfield  {author} {\bibinfo {author} {\bibfnamefont {T.}~\bibnamefont
  {Berlijn}}, \bibinfo {author} {\bibfnamefont {P.~C.}\ \bibnamefont
  {Snijders}}, \bibinfo {author} {\bibfnamefont {O.}~\bibnamefont {Delaire}},
  \bibinfo {author} {\bibfnamefont {H.~D.}\ \bibnamefont {Zhou}}, \bibinfo
  {author} {\bibfnamefont {T.~A.}\ \bibnamefont {Maier}}, \bibinfo {author}
  {\bibfnamefont {H.~B.}\ \bibnamefont {Cao}}, \bibinfo {author} {\bibfnamefont
  {S.~X.}\ \bibnamefont {Chi}}, \bibinfo {author} {\bibfnamefont
  {M.}~\bibnamefont {Matsuda}}, \bibinfo {author} {\bibfnamefont
  {Y.}~\bibnamefont {Wang}}, \bibinfo {author} {\bibfnamefont {M.~R.}\
  \bibnamefont {Koehler}}, \bibinfo {author} {\bibfnamefont {P.~R.}\
  \bibnamefont {Kent}}, \ and\ \bibinfo {author} {\bibfnamefont {H.~H.}\
  \bibnamefont {Weitering}},\ }\href {\doibase 10.1103/PhysRevLett.118.077201}
  {\bibfield  {journal} {\bibinfo  {journal} {Physical Review Letters}\
  }\textbf {\bibinfo {volume} {118}},\ \bibinfo {pages} {2} (\bibinfo {year}
  {2017})},\ \Eprint {http://arxiv.org/abs/1612.09589} {arXiv:1612.09589}
  \BibitemShut {NoStop}%
\end{thebibliography}
\end{document}